\documentclass[prd,twocolumn,aps,amsfonts,
floats,tightenlines,floatfix,superscriptaddress]{revtex4}
\usepackage[english]{babel}
\usepackage{epsfig}
\usepackage{psfrag}
\usepackage{graphicx}
\usepackage{amsmath}
\usepackage{amssymb}
\usepackage{dsfont}
\newcommand{\intq}{\int\!\!\frac{d^4q}{(2 \pi)^4}}
\newcommand{\intp}{\int\!\!\frac{d^4p}{(2 \pi)^4}}
\newcommand{\intpv}{\int\!\!\frac{d^3\vec{p}}{(2 \pi)^3}}
\newcommand{\intpE}{\int\!\frac{dp_{4}}{2 \pi}\,}

\newcommand{\phslash}{\hat{p}\!\!/\,}

\newcommand{\wpslash}{\omega_{p}\!\!\!\!\!\!/\,\,\,}
\newcommand{\pvslash}{\vec{p}\!\!/\,}

\begin{document}

\date{\today}

\title{Color-superconductivity\\ in the strong-coupling regime of Landau 
gauge QCD}
\author{D.~Nickel}
\affiliation{Institute for Nuclear Physics, Technical University Darmstadt, 
  Schlo{\ss}gartenstra{\ss}e 9, D-64289 Darmstadt, Germany}
\affiliation{Institute of Physics, University of Graz,
  Universit{\"a}tsplatz 5, A-8010 Graz, Austria}
\author{J.~Wambach}
\affiliation{Institute for Nuclear Physics, Technical University Darmstadt, 
  Schlo{\ss}gartenstra{\ss}e 9, D-64289 Darmstadt, Germany}
\affiliation{Gesellschaft f{\"u}r Schwerionenforschung mbH, Planckstra{\ss}e
  1, D-64291 Darmstadt, Germany}
\author{R. Alkofer}
\affiliation{Institute of Physics, University of Graz,
  Universit{\"a}tsplatz 5, A-8010 Graz, Austria}

\begin{abstract}

The chirally unbroken and the superconducting 2SC and CFL phases are
investigated in the chiral limit within a Dyson-Schwinger approach for the
quark propagator in QCD. The hierarchy of Green's functions is truncated such
that at vanishing density known results for the vacuum and at asymptotically
high densities the corresponding weak-coupling expressions are recovered.  The
anomalous dimensions of the gap functions are analytically calculated. Based
on the quark propagator the phase structure is studied, and results for the
gap functions, occupation numbers, coherence lengths and pressure differences
are given and compared with the corresponding expressions in the weak-coupling
regime. 
At moderate chemical potentials the quasiparticle pairing gaps are several
times larger than the extrapolated weak-coupling results.

\end{abstract}

\maketitle

\section{Introduction}
Strongly interacting matter as it existed in the early universe, resides in
the interior of compact stellar objects or is produced in heavy-ion collisions
is subject to extreme conditions. An understanding of the different phases of
such matter in terms of the fundamental degrees of freedom of QCD in the realm
of low temperatures and high quark densities is still missing although such
knowledge would be of fundamental interest. At sufficiently high density and
low temperatures strongly interacting matter is expected to be a color
superconductor, which has re-attracted a lot of interest in recent years (for
corresponding reviews see~\cite{reviews,Rischke:2003mt}). Due to asymptotic
freedom, it can be systematically studied at asymptotically large densities in
a weak coupling expansion~\cite{SchafHong}. At densities that are relevant for
the interior of neutron stars one is in the strongly coupled regime, and such
an expansion is lacking. Investigations~\cite{AlfRapp} are usually done within
Nambu-Jona--Lasino-type models in mean-field approximation. The main objective
of these studies is the identification of possible scenarios for the
color-superconducting state. Since the results are rather model dependent on a
quantitative level an approach, which is directly based on the QCD degrees of
freedom is highly desirable.

Furthermore, although it has been argued on general grounds that quarks become
deconfined at high densities, a corresponding detailed picture is missing. This
relates to the fact that already at vanishing temperatures and densities many
aspects of the confinement mechanisms remain elusive, in spite of recent
progress (see {\it e.g.\/} ref.~\cite{Greensite:2003bk}). Relations between
the confining field configurations and the Gribov horizon have been
established~\cite{Greensite:2004mh}. Those are in turn related to the infrared
behavior of QCD Green's functions, either in Coulomb
gauge~\cite{Zwanziger:2003de} or in Landau gauge~\cite{vonSmekal:2000pz}
QCD. In the Coulomb gauge quark confinement results already from an effective
one-gluon exchange picture~\cite{Zwanziger:2002sh,Nakamura:2005ux} and
confinement of colored composite states such as `diquarks' can be
studied~\cite{Alkofer:2005ug}. A detailed understanding of the infrared part
of the Yang-Mills sector is still lacking, however. On the other hand, in the
Landau gauge the gluon propagator has been determined quantitatively by
different methods with consistent results, see {\it e.g.\/}
\cite{Silva:2005hb,Sternbeck:2005tk,Fischer:2004uk,Pawlowski:2003hq,Alkofer:2003jj}
and references therein. The infrared behavior of the vacuum gluon and ghost
propagators, established by these and related studies, have provided the basis
for investigations of the quark propagator as well as extensions to finite
temperature for the pure Yang-Mills sector~\cite{Maas}. The results underline
non-trivial strong coupling of ghosts and gluons even at arbitrary high
temperatures.

It seems natural to suitably extend these methods to finite densities
especially for the strongly coupled regime, and compare to analytical results
obtained in the weak coupling limit. In the present work we proceed in this
direction by extending the truncation scheme of the pertinent Dyson-Schwinger
equations (DSE) proposed in ref.~\cite{Fischer:2003rp}. We implement momentum-
and energy-dependent self energies, which encode information about the
non-Fermi liquid behavior of the chirally unbroken phase at large quark
densities. This is important for the analysis of superconducting gap functions
already in the weakly coupled regime~\cite{Wang:2001aq}. We consider medium
effects on the quark-quark interaction in the random-phase approximation
similar to the Hard-Dense-Loop (HDL) approximation and thereby recover
the corresponding analytical results in the weakly coupled regime. The
presented results allow for a quantitative estimate for the validity of the
weak coupling results at a given chemical potential. This is the main
objective of the present investigation.

This paper is organized as follows: In sect.~II the truncated quark
Dyson-Schwinger equation  at non-vanishing chemical potential is derived.
In sect.~III the ultraviolet behavior of the gap functions is  extracted
analytically. In sect.~IV we provide expressions for occupation numbers,
diquark coherence lengths and the effective action. In sect.~V we discuss the
relevance and application of Luttinger's theorem for relativistic matter.
In sect.~VI we give numerical results for the unbroken phase. In sect.~VII
results for two superconducting phases (2SC and CFL) are presented, including
an estimate of the pressure from the Cornwall-Tomboulis-Jackiw (CJT)
action. Sect.~VIII closes with some concluding remarks and an outlook.

\section{QCD Green's functions}
\subsection{The quark propagator at non-vanishing chemical potential}

For the description of superconducting systems, it is advantageous to work
within the Nambu-Gor'kov formalism~\cite{Gorkov}. In the following we will only
consider the chiral limit and thus treat all flavors equally. Extending the
conventions and notations of refs.~\cite{Alkofer:2000wg,Roberts:2000aa}, the
unrenormalized QCD Lagrangian in Euclidean space at finite quark chemical
potential $\mu$ and vanishing quark masses reads
\begin{eqnarray}
  &&S^{(\mu)}\left[\psi,\bar{\psi},A_{\nu}^{a}\right]
  =
  S\left[\psi,\bar{\psi},A_{\nu}^{a}\right]+\bar \psi \gamma_4 \mu \psi
  \nonumber\\
  &=&
  \frac{1}{2}\bar{\Psi}\left(
    \begin{array}{cc}
      -\gamma_{\mu}D_{\mu}+\gamma_{4}\mu & 0\\
      0&  -\gamma_{\mu}D_{C\mu}-\gamma_{4}\mu
    \end{array}
  \right)\Psi\nonumber\\
  && \hspace{40mm}+
  \frac{1}{4}F^{a}_{\mu\nu}F^{a}_{\mu\nu}\,,
\end{eqnarray}
where $D_{C\mu}=\partial_{\mu}-igA^{T}_{\mu}$ corresponds to the 
charge-conjugate of the covariant derivative
$D_{\mu}=\partial_{\mu}+igA_{\mu}$. Using the charge-conjugation matrix of
Dirac spinors, $C=\gamma_{2}\gamma_{4}$,  the $8N_{c}N_{f}$-dimensional
bispinors $\Psi (x)$ are defined as
\begin{align}
  \Psi =
  \left(
    \begin{array}{c}
      \psi \\ \psi_{C} = C\bar{\psi}^{T}
    \end{array}
  \right)
  ,\quad&
  \bar{\Psi}=
  \left(\bar{\psi},\bar{\psi}_{C}=\psi^{T}C\right).
\end{align}
Note that the components of these bispinor field operators are not independent 
of each other~\cite{Rischke:2000ra}. In the following $N_c=3$ and $N_f=3$ will
be used.

In a covariant gauge, the renormalized quark DSE with appropriate
quark-wave-function and quark-gluon-vertex renormalization constants, $Z_{2}$
and $Z_{1F}$, respectively, is then given by
\begin{eqnarray}
\label{qDSE}
  \mathcal{S}^{-1}(p) &=& Z_{2}\mathcal{S}_{0}^{-1}(p)+ Z_{1F}\Sigma(p),
\end{eqnarray}
where 
\begin{widetext}
\begin{eqnarray}
 \mathcal{S}_{0}^{-1}(p) &=&
 \left(
   \begin{array}{cc}
     -i \vec{p}\cdot\vec{\gamma}-i(p_{4}+i\mu+gA_{4})\gamma_{4} & 0\\
     0 & -i \vec{p}\cdot\vec{\gamma}-i(p_{4}-i\mu-gA_{4}^{T})\gamma_{4}\\
   \end{array}
 \right)
\end{eqnarray}
\end{widetext}
is the inverse bare quark propagator in the presence of a static, isotropic
and homogeneous time component $A_4$ of the gluon field. The quark self energy
$\Sigma(p)$ is given by
\begin{eqnarray}
  \Sigma(p) = 
  -\intq\Gamma_{NG,\mu}^{(0)a}\mathcal{S}(q)
  \Gamma^{b}_{NG,\nu}(q,p)D^{ab}_{\mu\nu}(q-p).
\end{eqnarray}
Here $D_{\mu\nu}^{ab}(k)$ is the gluon propagator, and the bare and full
quark-gluon vertex in the Nambu-Gor'kov basis is defined as
\begin{align}
  \Gamma^{(0)a}_{NG\mu} =
  \frac{ig}{2}\left(
    \begin{array}{cc}
      \gamma_{\mu}\lambda^{a} & 0\\
      0 & -\gamma_{\mu}\lambda^{a T}
    \end{array}
  \right)
  \nonumber ,
\end{align}
\begin{align}  
  \Gamma^{a}_{NG\mu}(q,p) =
  \frac{ig}{2}\left(
    \begin{array}{cc}
      \Gamma^{a}_{\mu}(q,p) & \Delta^{a}_{C \mu}(q,p)\\
      \Delta^{a}_{\mu}(q,p) & \Gamma^{a}_{C \mu}(q,p) 
    \end{array}
  \right) , \label{NGvertex}
\end{align}
with $\Delta^{a}_{C \mu}(q,p)=-C\Delta^{a}_{\mu}(-p,-q)^{T}C$ and
$\Gamma^{a}_{C \mu}(q,p)=-C\Gamma^{a}_{\mu}(-p,-q)^{T}C$.
Flavor indices have been suppressed, they will be discussed in detail below. 

In a fixed gauge, the expectation value of $A_4$ is determined by its DSE,
{\it i.e.\/} by its equation of motion. In a covariant gauge this corresponds
to a vanishing expectation value of
\begin{eqnarray}
  \frac{\delta S^{(\mu)}[\psi,\bar{\psi},A_{\mu}^{a}]}{\delta A_{\mu}^{a}(x)}
  &=&
  -\left(\partial_{\nu}\delta^{ab}+gf^{abc}A^{c}_{\nu}(x)\right)F^{b}_{\nu\mu}(x)
  \nonumber \\
  &&-\frac{1}{2}\bar{\Psi}(x)\Gamma^{(0)a}_{NG\mu}\Psi(x)
  -\frac{1}{\lambda}\partial_{\mu}\partial_{\nu}A_{\nu}^{a}(x)
  \nonumber\\&&
  -igf^{abc}\partial_{\mu}\bar{c}_{b}(x)c_{c}(x),
\end{eqnarray}
where $c(x)$ and $\bar{c}(x)$ is the ghost and antighost field, respectively.
For a static, isotropic, homogenous, even-parity and $T$-symmetric ground
state $|\Omega\rangle$ in a fixed covariant gauge one has
\begin{eqnarray}
  \langle\Omega| D_{\nu}^{ab}F_{\nu\mu}^{b}(x)|\Omega\rangle
  &=& \nonumber\\
  -gf^{abc}\langle\Omega| A^{c}_{\nu}(x)F^{b}_{\nu\mu}(x)|\Omega\rangle &=&
  0
\end{eqnarray}
and, since the ghost propagator is real and symmetric,
\begin{eqnarray}
\langle\partial_{\mu}\bar{c}_{b}(x)c_{c}(x)\rangle=0~.
\end{eqnarray}
From this one eventually obtains
\begin{align}
  \rho^{a}(x) \propto
  \frac{1}{2}\mathrm{Tr}_{D,c,f,NG}\left(\mathcal{S}(x)\Gamma^{(0)a}_{NG
  4}\right) = 0~.
\end{align}
Therefore the static gluon fields ensure color neutrality ($\rho^{a}(x)=0$) and
can be implicitly determined by this condition~\cite{Gerhold:2003js,Dietrich:2003nu,Buballa:2005bv}.
For the symmetries stated above, this is equivalent to the exact solution of
the equations of motion.  For the quantities discussed in the following the
neutrality condition only slightly modifies the results, and thus it will be
neglected for simplicity.

\subsection{Gluon propagator and quark-gluon vertex} 

The quark propagator can be determined from eq.~(\ref{qDSE}) as a functional of
the gluon propagator and the quark-gluon vertex. In the Landau gauge, which
will be employed in the following, these Green's functions have been
investigated for the chirally broken phase at zero chemical potential by DSE
studies and by lattice calculations, see {\it
  e.g.\/}~\cite{Alkofer:2003jj,Fischer:2002hn,Sternbeck:2005tk,Silva:2005hb,Bowman:2004jm,Llanes-Estrada:2004jz,Skullerud:2004gp,Lin:2005zd}.
In the Landau gauge, the gluon propagator is parameterized by a single
dressing function $Z(k^{2})$,
\begin{eqnarray}
  D_{\mu\nu}^{ab}(k^{2}) &=& \delta_{ab}
  \left( \delta_{\mu\nu} - \frac{k_\mu k_\nu}{k^2} \right) 
  \frac{Z(k^{2})}{k^2}.
\end{eqnarray}

The quark-gluon vertex possesses in general twelve linearly independent Dirac
tensor structures (see {\it e.g.\/} ref.\ \cite{Alkofer:2000wg}) with the 
vector-type coupling $\propto \gamma_{\mu}$ being the only one present at tree
level. It has been shown that some of the other tensor structures are
qualitatively important when studying the quark
DSE~\cite{Fischer:2003rp,Alkofer:2003jj}. For simplicity, in this exploratory
study the quark-gluon vertex is assumed to be of the form
\begin{eqnarray}
  \Gamma^{a}_{\mu}(p,q) &=&
  ig\Gamma((p-q)^{2})\gamma_{\mu}\frac{\lambda^{a}}{2}.
\label{AbelVertex}  
\end{eqnarray}
The unknown function $\Gamma(k^{2})$ has been chosen such that the quark
propagator is multiplicatively renormalizable and agrees with perturbation
theory in the ultraviolet~\cite{Fischer:2003rp}. It can also be determined
from quenched lattice results of the quark and gluon
propagator~\cite{Bhagwat:2003vw} or finite-volume results for the
corresponding DSE's in combination with lattice data for the quark
propagator~\cite{Fischer:2005nf}. Beyond the vector-type coupling contained in
(\ref{AbelVertex}) at least the  term proportional to $p_\mu+q_\mu$ is of
qualitative significance~\cite{Fischer:2003rp,Alkofer:2003jj}. In the chiral
limit such a scalar-type term is, however, only non-vanishing in the chirally
broken phase because such a contribution violates chiral symmetry. As we are
interested here in superconducting phases this scalar contribution will not be
present, at least, in the chiral limit. Therefore, the use of the
approximation (\ref{AbelVertex}) is certainly sufficient for an exploratory
study.

In the following only the product
\begin{eqnarray}
  \alpha_{s}(k^{2}) &=& \frac{Z_{1F}}{Z_{2}^{2}}
  \frac{g^{2}}{4\pi}Z(k^{2})\Gamma(k^{2}),
\end{eqnarray}
will enter the quark propagator DSE. Here $Z_{1F}$ is the quark-gluon coupling
and $Z_2$ the quark-wave-function renormalization constant, respectively.
We will refer to $\alpha_{s}(k^{2})$ as the effective strong running coupling
because, especially in the framework of DSEs, it is a possible non-perturbative
extension of the coupling into the infrared. 

It is, of course, expected that the gluon propagator and the quark-gluon
vertex  undergo changes when a non-vanishing chemical potential is
introduced. Note that in superconducting phases anomalous vertices, the
$\Delta^{a}_{\mu}$ in eq.\ (\ref{NGvertex}), are induced. These are neglected
in the study reported in this paper. Thus, taking into account the matrix
structure implied by the  Nambu-Gor'kov formalism and the approximation
(\ref{AbelVertex}) we will use
\begin{eqnarray}
  \Gamma^{a}_{NG\mu}(p,q) &=& \Gamma((p-q)^2)\Gamma^{(0)a}_{NG\mu},
\label{NGAbelVertex}  
\end{eqnarray}
with $\Gamma(k^{2})$ taken from quenched vacuum studies.

Since we are primarily interested in chirally unbroken phases at non-vanishing
chemical potential, it is important to incorporate medium effects like damping
and screening by particle-hole excitations. In this paper we add the in-medium
polarization tensor with 'bare' quark propagators to the inverse gluon
propagator. This is not self-consistent and needs to be investigated in further
studies. Nevertheless, the resulting quark DSE turns out to be a generalization
of the HDL approximation, with the important difference that  the infrared
behavior of the gluon propagator and quark-gluon vertex is
nontrivial. Furthermore, even in the superconducting phases, the assumption of
bare quark propagators will {\it a posteriori} turn out to be much better
suited than employing the vacuum quark propagator, {\it i.e.\/} the one
reflecting dynamical chiral symmetry breaking.

\begin{widetext}
The renormalized medium polarization tensor is generically given by
\begin{eqnarray}
  \Pi_{\phantom{{\rm med} \,}\mu\nu}^{{\rm med} \, ab}(p)=\frac{1}{2}Z_{1F}
  \intq
  \mathrm{Tr}_{D,c,f,NG}\left( \left[\Gamma^{(0)a}_{NG\mu}\mathcal{S}(q)
    \Gamma_{NG\nu}^{b}(q,p-q)\mathcal{S}(p-q)\right]
  - \left[ \ldots \right] _{\mu =0} \right)~.
\end{eqnarray}
For the first part we employ the approximation of vanishing quark
self-energies and the vertex (\ref{NGAbelVertex}). Then this expression can be
straightforwardly reduced to
\begin{eqnarray}
  \Pi_{\phantom{{\rm med} \,}\mu\nu}^{{\rm med} \, ab}(p)
  &=&
  -\frac{Z_{1F}}{Z_{2}^{2}}\frac{g^{2}N_{f}}{2}\delta^{ab}\Gamma(p^2)
  \intq\mathrm{Tr}_{D}\left( 
  \left[\gamma_{\mu}S_{0}(q)\gamma_{\nu}S_{0}(p-q)\right] - 
  \left[ \ldots \right] _{\mu =0} \right)
  \nonumber\\&=&
  -\delta^{ab}\frac{2\pi N_{f}\alpha_{s}(p^{2})}{Z(p^2)}
  \intq\mathrm{Tr}_{D}\left(
  \left[\gamma_{\mu}S_{0}(q)\gamma_{\nu}S_{0}(p-q)\right]
  - \left[ \ldots \right] _{\mu =0} \right).
\end{eqnarray}
\end{widetext}
With help of projectors transverse and longitudinal to the
medium~\cite{Kapusta:1989tk}, $$P^T_{4 4}=P^T_{i 4}=P^T_{4 i}=0,  \quad P^T_{i
  j}=\delta_{ij}-\frac{p_i p_j}{\vec p^2} ,$$ and 
$$P^{L}_{\mu\nu}= (\delta_{\mu\nu}-p_\mu p_\nu /p^2) - P^{T}_{\mu\nu} ,$$  
respectively, this polarization tensor can be expressed by two functions, 
$G(|\vec{p}|,p_{4})$ and $F(|\vec{p}|,p_{4})$,
\begin{eqnarray}
  Z(p^2)\Pi_{\mu\nu}^{ab}(p)
  &=&
  G(p) \delta^{ab}P^{T}_{\mu\nu}+  F(p) \delta^{ab}P^{L}_{\mu\nu}.
  \nonumber \\
\end{eqnarray}
The evaluation of this functions is well known perturbatively, and in the
present case only the coupling is replaced by the running coupling. For small
external momenta, the result is
\begin{eqnarray}
  G(|\vec{p}|,p_{4}) &=&
  m^{2}(p^{2})\frac{ip_{4}}{|\vec{p}|}  
  \left[
    \left(
      1-\left(\frac{ip_{4}}{|\vec{p}|}\right)^{2}
    \right) \right.
\nonumber \\    
   && \left. Q\left(\frac{ip_{4}}{|\vec{p}|}
    \right)+\frac{ip_{4}}{|\vec{p}|}
  \right],
  \\  
  F(|\vec{p}|,p_{4}) &=& 2 m^{2}(p^{2}) \frac{p_{4}^{2}+\vec{p}^{2}}{\vec{p}^{2}}
 \nonumber \\  
  &&\left[
    1-\frac{ip_{4}}{|\vec{p}|}Q\left(\frac{ip_{4}}{|\vec{p}|}\right)
  \right],
  \label{GF}
\end{eqnarray}
with $Q(x) = \frac{1}{2}\ln \frac{x+1}{x-1}$
and 
\begin{eqnarray}
m^{2}(p^{2})={N_{f}\alpha_{s}(p^{2})\mu^{2}}/{\pi}.
\label{m2}
\end{eqnarray}
Adding the medium polarization to the inverse gluon propagator leads to
\begin{eqnarray}
  D_{\mu\nu}^{ab}(p)
  &\approx&
  \left(D^{vac\,-1}_{\mu\nu}(p^{2}) + 
  \Pi_{\phantom{{\rm med} \,}\mu\nu}^{{\rm med} \, ab}(p)\right)^{-1}
  \nonumber \\
  &=&
  \delta^{ab}
  \Bigl(\frac{p^{2}}{p^{2}+G(p)} P^{T}_{\mu\nu}
  \nonumber \\
  &&+\frac{p^{2}}{p^{2}+F(p)}
  P^{L}_{\mu\nu}\Bigr)
\frac{Z(p^2)}{p^{2}}~.
\end{eqnarray}
Taking into account this form of the in-medium polarization, Debye screening
and Landau  damping are included, similar as in the HDL approximation. As it
is phrased here, it becomes evident that these phenomena have a
non-perturbative origin and require the knowledge of the infrared behavior of
Schwinger functions, in particular of $\alpha_{s}(k^{2})$.

\subsection{The truncated quark DSE at non-vanishing chemical potential}
\begin{figure}
\hspace{-.5cm}\includegraphics[width=8cm]{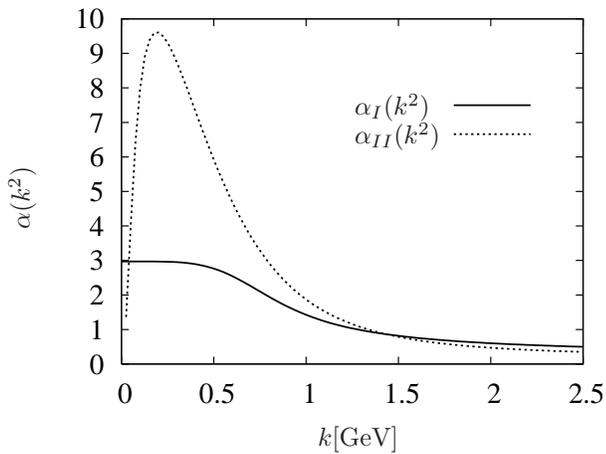}
\caption{Two versions of the strong running coupling in the numerical solution of the DSE (\ref{gap}).  The full line represents the coupling $\alpha_{I}(k^{2})$ extracted from the coupled gluon, ghost and quark DSE's~\cite{Fischer:2003rp} and the dashed line $\alpha_{II}(k^{2})$ extracted
from quenched lattice data together with an abelian approximation for the
quark-gluon vertex~\cite{Bhagwat:2003vw}.}
\label{salpha}
\end{figure}

Summarizing the above considerations we arrive at a truncated, self-consistent
DSE for the quark propagator. Treating the components in the
Nambu-Gor'kov basis separately we define~\cite{Rischke:2003mt}
\begin{eqnarray}
\mathcal{S}_{0}(p)&=& 
\left(
  \begin{array}{cc}
    S_{0}^{+}(p) & 0\\
    0 & S_{0}^{-}(p) 
  \end{array}
\right), \nonumber \\ \nonumber \\ 
\mathcal{S}(p)&=&
\left(
  \begin{array}{cc}
    S^{+}(p)& T^{-}(p) \\
    T^{+}(p)& S^{-}(p) 
  \end{array}
\right), \nonumber \\ \nonumber \\ 
\Sigma(p)&=&
\left(
  \begin{array}{cc}
    \Sigma^{+}(p)&\Phi^{-}(p)\\
    \Phi^{+}(p)&\Sigma^{-}(p)
  \end{array}
\right),
\end{eqnarray}
with
\begin{eqnarray}
A^{-}(p) &=& -C(A^{+}(-p))^{T}C 
\\ 
{\rm for} \quad A^\pm &\in & \{ 
S_{0}^\pm , S^\pm , T^\pm
\},  \nonumber 
\end{eqnarray}
$\Sigma^{+}(p)+\Sigma^{-}(p) =
\gamma_{4}\left(\Sigma^{+}(p)+\Sigma^{-}(p)\right)^{\dagger}\gamma_{4}$ and
\begin{eqnarray}
\Phi^{-}(p)   = \gamma_{4}(\Phi^{+}(p))^{\dagger}  \gamma_{4}
\end{eqnarray}
for a real action. One then obtains~\cite{Manuel:2000nh}
\begin{eqnarray}
  T^{\pm} &=&
  -Z_{1F}\left(Z_{2}{S^{\mp}_{0}}^{-1}+Z_{1F}\Sigma^{\mp}\right)^{-1}
  \Phi^{\pm}S^{\pm},\nonumber \\
  {S^{\pm}}^{-1} &=& \hphantom{-}Z_{2}{S^{\pm}_{0}}^{-1}+Z_{1F}\Sigma^{\pm}
  \nonumber \\ &-&
  Z_{1F}^{2}\Phi^{\mp}\left(Z_{2}{S^{\mp}_{0}}^{-1}+Z_{1F}\Sigma^{\mp}
  \right)^{-1}\Phi^{\pm},
  \nonumber \\ 
\end{eqnarray}
combined with
\begin{eqnarray}
  \Sigma^{+}(p) &=& \hphantom{-}\frac{Z_{2}^{2}}{Z_{1F}} \pi \intq
  \gamma_{\mu} \lambda_{a} S^{+}(q) \gamma_{\nu} \lambda_{a}
  \nonumber \\ &&
  \left(\frac{\alpha_{s}(k^{2})P^{T}_{\mu\nu}}{k^{2}+G(k)}
  +\frac{\alpha_{s}(k^{2})P^{L}_{\mu\nu}}{k^{2}+F(k)}\right) ,
  \nonumber \\
  \Phi^{+}(p) &=& -\frac{Z_{2}^{2}}{Z_{1F}} \pi \intq
  \gamma_{\mu} \lambda_{a}^{T} T^{+}(q) \gamma_{\nu} \lambda_{a}
  \nonumber \\ &&
  \left(\frac{\alpha_{s}(k^{2})P^{T}_{\mu\nu}}{k^{2}+G(k)}+
    \frac{\alpha_{s}(k^{2})P^{L}_{\mu\nu}}{k^{2}+F(k)}
  \right). \nonumber \\ 
  \label{gap}
\end{eqnarray}
Note that the running coupling $\alpha_{s}(k^{2})$ also enters via the
functions $G(|\vec{k}|,k_{4})$ and $F(|\vec{k}|,k_{4})$. On the other hand,
this running coupling is the only input into eqs.\ (\ref{gap}). In the
following we will present results for two different couplings.
Both couplings are derived in a MOM regularization scheme.
One will be the  running coupling determined from the coupled gluon, ghost and
quark DSEs~\cite{Fischer:2003rp}, denoted by $\alpha_{I}(k^{2})$ in the
following, where
\begin{eqnarray}
  \alpha_{I}(k^{2}) &=&
  \frac{\alpha_{I}(0)}{\ln\left(
      e+
      a_{1}\left(k^{2}/\Lambda_{I}^{2}\right)^{a_{2}}+
      b_{1}\left(k^{2}/\Lambda_{I}^{2}\right)^{b_{2}}
    \right)},\quad
\end{eqnarray}
with $\alpha_{I}(0)=2.972$, $a_{1}=1.106$, $a_{2}=2.324$, $b_{1}=0.004$,
$b_{2}=3.169$ and $\Lambda_{I}=0.714\mathrm{GeV}$. As the quark-gluon vertex used in this study contains a sizeable ``scalar'' contribution (which is due to dynamical chiral symmetry breaking) the corresponding running coupling together with the approximation (\ref{AbelVertex}) underestimates the chiral condensate
significantly but should be, as mentioned above, the appropriate choice in chirally restored phases. We consider this coupling as a lower bound. On the other hand, the running coupling extracted from quenched lattice data together with an abelian approximation for the quark-gluon vertex~\cite{Bhagwat:2003vw} serves as an upper bound and will be denoted by $\alpha_{II}(k^{2})$ in the following. As the corresponding parameterization is quite lengthy, we refer to
eqs. (8)-(16) of ref.~\cite{Bhagwat:2003vw} for its definition. Both couplings are shown in Fig.~\ref{salpha}.

In order to obtain a self-consistent solution of the system of equations (\ref{gap}) with (\ref{GF}), (\ref{m2}) and the running coupling as input, we
will first consider the color and flavor structure of the quark self energy. 
We will restrict ourselves to scalar pairing and treat the 2-flavor color-superconducting (2SC) and color-flavor locked (CFL) phases separately.  In the chiral limit, the pairing structure can be described by a symmetric matrix $M$ such that the projectors on its eigenspaces $P_{i}$ together with the matrices  $M_{i}:= M P_{i}$ form a closed basis under the transformations
$P_{i} \to \lambda_{a}P_{i}\lambda_{a}$  and $M_{i} \to \lambda^{T}_{a}M_{i}
\lambda_{a}$, respectively. Therefore, we can parameterize the r.h.s.\ of eqs.\ (\ref{gap}) by the renormalization point independent, Dirac-algebra valued self energies $\Sigma^{+}_{i}(p)$ and $\phi^{+}_{i}(p)$:
\begin{eqnarray}
  \Sigma^{+}(p)
  &=&
  \frac{Z_{2}}{Z_{1F}}
  \sum_{i} \Sigma^{+}_{i}(p) P_{i},\\
  \Phi^{+}(p)
  &=&
  \frac{Z_{2}}{Z_{1F}}
  \sum_{i} \phi^{+}_{i}(p) M_{i}.
\end{eqnarray}
For an even-parity and $T$-symmetric phase
those are given by~\cite{Pisarski:1999av}
\begin{eqnarray}
  \Sigma^{+}_{i}(p) &=&
  -i\pvslash\Sigma^{+}_{A,i}(p)-i\wpslash\Sigma^{+}_{C,i}(p)
 \nonumber  \\ 
 & = &
  \gamma_{4}\sum_{e=\pm1}\Sigma_{e,i}^{+}(p)\Lambda^{e}_{\vec{p}},
   \\
  \phi^{+}_{i}(p) &=&
  \left(
    \gamma_{4}\phslash\phi^{+}_{A,i}(p)
    +\phi^{+}_{C,i}(p)
  \right)\gamma_{5}
 \nonumber  \\ 
  &=&
  \gamma_{5}\sum_{e=\pm1}\phi_{e,i}^{+}(p)\Lambda^{e}_{\vec{p}},  
\end{eqnarray}  
where we made use of the positive and negative energy projectors
$\Lambda^{\pm}_{\vec{p}}=\frac{1}{2}\left(1\pm i\gamma_{4}\phslash\right)$,
$\pvslash =\vec{p}\cdot\vec{\gamma}$, $\hat{p}=\vec{p}/\vert \vec{p}\vert$,
$\wpslash = \omega_{p}\gamma_{4}$ and $\omega_{p}=ip_{4}+\mu$.

Introducing positive parameters $\delta_{i}>0$ via the decomposition 
$M^{\dagger}M = \sum_{i}\delta_{i}P_{i}$ the quark particle-particle
propagator can be written as
\begin{widetext}
\begin{eqnarray}
  Z_{2} S^{+}(p)
  =  -\gamma_{4} \sum_{i,e=\pm} P_{i} \Lambda^{-e} 
  \frac{
    \left(-i p_{4}-\mu\right)\left(1+{\Sigma_{C,i}^{+}(p)}^{*}\right)
    +e|\vec{p}|\left(1+{\Sigma_{A,i}^{+}(p)}^{*}\right)
  }{
    |(ip_{4}-\mu)(1+\Sigma_{C,i}^{+}(p))+e|\vec{p}|(1+\Sigma_{A,i}^{+}(p))|^{2}
    +\delta_{i}|\phi_{e,i}(p)|^{2}
  }, 
  \nonumber\\ 
\end{eqnarray}
\end{widetext}
where we have made use of the relations $\Sigma_{F,i}(p)={\Sigma_{F,i}(-p)}^{*}$.
The zero of the numerator at $p_{4}=0$ defines the quasiparticle Fermi momenta 
$p_{F}$, and the zero in the denominator provides the corresponding dispersion 
relation. To first approximation, the energy gap $\Delta_{i}^{e}$ in the excitation spectrum is therefore given by
\begin{eqnarray}
  \label{DeltaE}
  \Delta_{i}^{e}
  &\simeq&
  \left|
    \frac{\sqrt{\delta_{i}}\,\phi_{e,i}(p)}{1+\Sigma_{C,i}^{+}(p)}
  \right|_{|\vec{p}|=p_{F},p_{4}=0}.
\end{eqnarray}

\section{Ultraviolet finiteness of the gap functions}
Similar to the ultraviolet analysis of the quark mass function in the chirally
broken phase~\cite{Gusynin:1986fu} we determine here the ultraviolet behavior 
of the gap functions. For large external momenta $p$, such that 
$\phi^2 (p) \ll p^2$,  $\mu^2\ll p^2$ and $m^2\ll p^2$, the breaking of Lorentz 
covariance is negligibly small, and thus the self-energies 
$\Sigma^{\pm}(p)$ and $\Phi^{\pm}(p)$ are to a very good approximation functions of the four-momentum squared, $p^2$, only. Furthermore, in the denominators of the integral kernels the self-energies can be safely neglected. 
Also the medium modifications of the gluon propagator will then be
insignificant, and the gap equations reduce to
\begin{eqnarray}
 && \sum_{i}\phi^{+}_{C,i}(p)M_{i}
  \simeq
  \nonumber \\ && \hspace{.5cm}
  - 3\pi \sum_{i}
  \intq 
   \frac{\phi^{+}_{C,i}(q)}{q^{2}}
  \frac{\alpha_{s}(k^{2})}{k^{2}}
  \lambda_{a}^{T}
  M_{i}
  \lambda_{a},
  \nonumber \\ \\
  &&\sum_{i}\phi^{+}_{A,i}(p)M_{i}
  \simeq
  \nonumber \\ && \hspace{.5cm}
  -\sum_{i}
  \pi \intq
  \frac{\phi^{+}_{A,i}(q)}{q^{2}}
  \frac{\alpha_{s}(k^{2})}{k^{2}}
  \lambda_{a}^{T}
  M_{i}
  \lambda_{a}
  \nonumber \\ && \hspace{.5cm}\times
  \left(
    \hat{p}\cdot\hat{q}
    \left(1-2\frac{k_{4}^{2}+(\hat{q}\cdot\vec{k})^{2}}{k^{2}}\right)
  \right)
  .
\end{eqnarray}
As the running coupling is a slowly varying function for large momenta
it is safe to apply the angular approximation $\alpha_{s}(k^{2}) 
\approx \alpha_{s}(p^2\,\theta(p^{2}-q^{2}) + q^2\,\theta(q^{2}-p^{2}))$. 
The remaining angular integrations can then be done analytically.
The gap function $\phi^{+}_{A,i}(p)$ decreases for large $p^2$ at least like
$1/p^4$ times logarithmic corrections, and thus to order $1/p^2$, one has
$\phi^{+}_{A,i}(p)\approx 0$. The equation for the gap function
$\phi^{+}_{C,i}(p)$ then reads
\begin{eqnarray}
  \sum_{i}\phi^{+}_{C,i}(p)M_{i}
  &\simeq&
  -\frac{3}{16\pi}
  \sum_{i}
  \lambda_{a}^{T}
  M_{i}
  \lambda_{a}
  \nonumber\\&&\hspace{3mm}
  \left(
  \frac{\alpha_{s}(p^{2})}{p^{2}}
  \int^{p^{2}} \!dq^{2}\,
  \phi^{+}_{C,i}(q) \right.
  \nonumber\\ &&\hspace{3mm} \left.
  +
  \int_{p^{2}} \!dq^{2}\,
  \frac{\alpha_{s}(q^{2})\phi^{+}_{C,i}(q)}{q^{2}}
  \right).\nonumber\\ \label{UV}
\end{eqnarray}

The color-antitriplet channel is attractive and for this channel we get
$\lambda_{a}^{T}M_{i}\lambda_{a}=-\frac{8}{3}M_{i}$ as compared to
$\lambda_{a}\lambda_{a}= \frac{16}{3}=4C_F$. The color sextet channel is
repulsive. For this one we have $\lambda_{a}^{T}M_{i}\lambda_{a}
= \frac{4}{3}M_{i}$. The anomalous dimensions of the gap functions
$\gamma_{\phi,i}$, to one-loop order can then be read off from the coefficient
in eq.\ (\ref{UV}). Comparing to the corresponding anomalous dimension
of the mass function $\gamma_{m}=12/(33-2N_{f})$, they are given by
\begin{eqnarray}
\gamma_{\phi,\bar 3} = \phantom{-}\gamma_{m}/2 =\phantom{-} 6/(33-2N_{f})
\end{eqnarray} 
and 
\begin{eqnarray}
\gamma_{\phi,6} = -\gamma_{m}/4 = -3/(33-2N_{f})
\end{eqnarray}  
in attractive and repulsive channels, respectively. Similar to the chiral quark
condensate, see {\it e.g.\/} ref.~\cite{Roberts:1994dr}, one is now able to
define a renormalization-group independent diquark condensate. The asymptotic
behavior of $\phi^{+}_{C,i}(p)$ is given by the so-called 'regular form'
\begin{eqnarray}
  \phi^{+}_{C,i}(p)
  &\propto&
  \frac{1}{p^{2}}\left(\ln\left(\frac{p^{2}}{\Lambda^{2}}\right)\right)
  ^{\gamma_{\phi,i}-1}. 
\end{eqnarray}

\section{Relevant quantities}
\subsection{Occupation numbers and diquark correlations}
\label{occcor}
Once the quark propagator is known, one can extract number densities,
occupation numbers and the diquark coherence lengths. Within the Euclidean
formalism, the number density $\rho$ is calculated as the derivative of the
generating functional of the connected Green's functions with respect to the
chemical potential $\mu$. For the homogeneous phases, considered here it is
given by
\begin{eqnarray}
  \label{densityrho}
  \rho &=& Z_{2}\langle\langle \psi^{\dagger}(\vec{x})\psi(\vec{x})
  \rangle\rangle
  \nonumber\\&=&
  \lim_{x_{4}\rightarrow i 0^{+}} 
  \frac{1}{2}\mathrm{Tr}_{D,c,f,NG}
  \left(
    Z_{2}\,\gamma_{4}\otimes\mathds{1}_{NG}\mathcal{S}(x_{4},0)
  \right)
  \nonumber\\&=&
  \intpv\intpE \mathrm{Tr}_{D,c,f}
  \left(Z_{2}\,\gamma_{4}\otimes\mathds{1}_{NG}S^{+}_{q}(p)\right)
  \nonumber\\&=&
  \sum_{i}\frac{g_{i}}{(2\pi)^{3}}\int d^{3}p \,\,n_{i}(p) ,
\end{eqnarray}
where $g_{i}=2\, \mathrm{rank}(P_{i})$ is a degeneracy factor, and the
occupation numbers $n_{i}(p)$ read
\begin{eqnarray}
  n_{i}(p) &=& \frac{Z_{2}}{4\pi}\int_{-\infty}^{\infty}\!\!dp_{4}\,\,
  \mathrm{Tr}_{D}\left(\gamma_{4}S^{+}_{i}(p_{4},p)\right)~.
  \nonumber\\
\end{eqnarray}
The $p_{4}$-integration has to performed first to make this expression
well-defined. The relation between the density and the Fermi momentum will be
further discussed in sect.~\ref{app2}.

The quark-quark correlation lengths provides a measure of the size of the
paired diquarks. They can be determined from the anomalous propagator
\begin{eqnarray}
  T^{+}_{i}(x-y)
  &=&
  \langle\langle \psi(x)^{T} C M_{i} \psi(y)\rangle\rangle
  \nonumber\\&=&
  \intp e^{ip(x-y)}\,M_{i} \sum_{e=\pm}
  T^{+}_{i,e}(p)\,\Lambda^{e}_{\vec{p}}~.
 \nonumber\\
\end{eqnarray}
For a given pairing pattern, $M_{i}$, the coherence length $\xi_{i,e}$ is
defined as
\begin{eqnarray}
  \xi_{i,e} &=&
  \frac{
    \int\! d^{3}x\,\, |\vec{x}|^{2} |T^{+}_{i,e}(0,\vec{x})|^{2}
  }{
    \int\! d^{3}x\,\, |T^{+}_{i,e}(0,\vec{x})|^{2}
  }
  \nonumber\\&=&
  \frac{
    \int\! d^{4}p\,\, |\nabla_{\vec{p}} T^{+}_{i,e}(p)|^{2}
  }{
    \int\! d^{4}p\,\, |T^{+}_{i,e}(p)|^{2}
  }.
\end{eqnarray}

\subsection{The effective action}
Studying different phases,  the question naturally arises which phase is
energetically preferred. To study this, we estimate the corresponding pressure
difference by employing the Cornwall-Jackiw-Tomboulis (CJT)
formalism~\cite{Cornwall:1974vz}, which provides the effective action $\Gamma$
as a functional of the expectation values of fields and propagators in
presence of local and bilocal source terms. In particular, for QCD in the
Nambu-Gor'kov formalism the functional dependence on the quark propagator is
given by~\cite{Rischke:2003mt,Schmitt:2004et}
\begin{eqnarray}
  \label{CJT}
  \Gamma[\mathcal{S}] =& 
  -&\frac{1}{2}\mathrm{Tr}_{p,D,c,f,NG}\mathrm{Ln}\mathcal{S}^{-1}
  \nonumber \\ &+&
  \frac{1}{2}\mathrm{Tr}_{p,D,c,f,NG}
  \left(1-Z_{2}\mathcal{S}_{0}^{-1}\mathcal{S}\right)+
  \Gamma_{2}[\mathcal{S}] .
  \nonumber \\ 
\end{eqnarray}
Here $\Gamma_{2}[\mathcal{S}] $ is the 2-particle irreducible part of this 
effective action. Note that the self energy can be expressed as the functional
derivative with respect to $\Gamma_{2}[\mathcal{S}] $, 
$Z_{1F}\Sigma[\mathcal{S}]= -2{\delta\Gamma_{2}[\mathcal{S}]}/{\delta
\mathcal{S}}$, and thus possesses a corresponding functional dependence on the
full quark propagator. One can ``integrate'' the DSE and obtains an
approximate, yet thermodynamically consistent, effective action: 
\begin{eqnarray} \Gamma_{2}[\mathcal{S}] &\simeq&
-\frac{1}{4}\mathrm{Tr}_{p,D,c,f,NG}(1-Z_{2}\mathcal{S}_{0}^{-1}\mathcal{S})
\nonumber \\ &+& {\mathrm const.} \, .
\end{eqnarray}

\section{On Luttinger's theorem}
\label{app2}
Luttinger's theorem can be summarized as follows: Provided the fermion
propagator is positive at the Fermi energy, $p_4=0$, the volume of the
Fermi surface at fixed density is independent of the interaction. The proof of
this theorem is based on the fact that the functional
$\Gamma_{2}[\mathcal{S}]$ is invariant under shifts in the momentum, {\it
  i.e.}, 
\begin{eqnarray}
  \delta\Gamma_{2}[\mathcal{S}] =
  -\frac{1}{2}Z_{1F}\mathrm{Tr}_{p,D,c,f,NG}\left[
    \Sigma[\mathcal{S}]\frac{\partial}{\partial p_{4}}\mathcal{S}
  \right]
  =0.
\end{eqnarray}
Using
$Z_{2}\,\gamma_{4}\otimes\mathds{1}_{NG}=
i\frac{\partial}{\partial p_{4}}\left(\mathcal{S}^{-1}-Z_{1F}\Sigma\right)$, 
eq.~(\ref{DeltaE}) and the spectral representation of the propagator one can
show that
\begin{eqnarray}
  \rho =
  \frac{1}{2}\intpv
  \left(
    \left.
      \frac{i}{2\pi}\mathrm{Tr}_{D,c,f,NG}\mathrm{Log}\left(\mathcal{S}^{-1}
      \right)
    \right|_{p_{4}=0^{-}}
  \right. &&
  \nonumber\\
  \left.
    \left.
      -\frac{i}{2\pi}\mathrm{Tr}_{D,c,f,NG}\mathrm{Log}\left(\mathcal{S}^{-1}
      \right)
    \right|_{p_{4}=0^{+}}
  \right) &&
\end{eqnarray}
where the cut in the complex logarithm has been, as usual, put onto the
negative real half-axis. This, for a single fermion species, amounts to
\begin{eqnarray}
  &&
  \left.
    \frac{i}{2\pi}\mathrm{Tr}_{D}\mathrm{Log}\left(S^{-1}\right)
  \right|_{p_{4}=0^{-}}
  \left.
    -\frac{i}{2\pi}\mathrm{Tr}_{D}\mathrm{Log}\left(S^{-1}\right)
  \right|_{p_{4}=0^{+}}
  \nonumber\\&=&
  \frac{i}{\pi}\ln
  \left.\left(
    i\omega_{p}C+\sqrt{A^{2}p^{2}+B^{2}}
  \right)\right|_{p_{4}=0^{-}}
  \nonumber\\&&
  -\left.\left(
    i\omega_{p}C+\sqrt{A^{2}p^{2}+B^{2}}
  \right)\right|_{p_{4}=0^{+}}
  \nonumber\\&=&
  \left\{
    \begin{array}{lr}
      2 \quad {\rm if}  & D(\vec{p},p_{4}=0)<0\\
      0 \quad {\rm if}   & D(\vec{p},p_{4}=0)>0
    \end{array}
  \right.,
\end{eqnarray}
where $D=\vert\vec{p}\vert^{2}A^{2}+\omega_{p}^{2}C^{2}+B^{2}$. Since the
Fermi surface in a Fermi liquid is defined by $D(\vec{p}=p_{F},p_{4}=0)=0$, it
is natural to extend this definition to a sign change in
$D(\vec{p},p_{4}=0)$. For a gapped mode this corresponds to a
singularity. Finally, we note that
\begin{eqnarray}
  \mathrm{Tr}_{NG}\mathrm{Log}\left(\mathcal{S}^{-1}\right)
  &=&
  \mathrm{Log}\left(
    S^{+\,-1}
    \left(
      Z_{2}S_{0}^{-\,-1}+Z_{1F}\Sigma^{-}
    \right)
  \right),
  \nonumber \\
\end{eqnarray}
as well as $D(\vec{p},p_{4}=0)$ of $S^{+}$ and
$\left(Z_{2}S_{0}^{-\,-1}+Z_{1F}\Sigma^{-}\right)^{-1}$ change sign at the
same momenta,
since $S^{+}$ is only invertible iff 
$Z_{2}(S_{0}^{-})^{-1}+Z_{1F}\Sigma^{-}$ is.
In an isotropic phase one therefore concludes that
\begin{eqnarray}
  \rho &=&
  \frac{1}{3\pi^{2}}\sum_{i}p_{F,i}^{3} .
\end{eqnarray}
Especially, the Fermi momenta $p_{F,i}$ are calculated by the zeros and 
poles in $\det_{D,c,f}\left(S^{+}\right)|_{p_{4}=0}$. We can therefore easily
determine the density as long as the symmetry
$\Sigma_{F,i}(p)={\Sigma_{F,i}(-p)}^{*}$ in our numerical solutions is not
spontaneously broken.

\section{Results for the unbroken phase}
Before turning to the superconducting phases and in order to provide a
case for comparison,  the unbroken phase in the chiral limit is studied
first. In eqs.\ (\ref{gap}) the functions $\Phi^{\pm}$, and accordingly
$T^\pm$, are set to zero. This yields a self-consistent integral equation for
the self energy $\Sigma^+$, which we solve numerically.

To understand the behavior of this numerical solution, we derive an approximate
form for its imaginary part. First, the inverse propagator is decomposed into
positive and negative energy parts
\begin{eqnarray}
  S^{+}(p)^{-1}
  &=&
  \gamma_{4}\Lambda_{\vec{p}}^{+}S^{+}_{+}(p)^{-1}+
  \gamma_{4}\Lambda_{\vec{p}}^{-}S^{+}_{-}(p)^{-1}~.
\end{eqnarray}
The quasiparticle propagator
\begin{eqnarray}
  S^{+}_{+}(p)
  &=&
  Z_{2}^{-1}\left(-ip_{4}+\mu-|\vec{p}|+\Sigma^{+}_{+}(p)\right)^{-1}
  \nonumber\\&=&
  Z_{2}^{-1}\Biggl(
    -ip_{4}\left(1-\frac{\mathrm{Im}\Sigma^{+}_{+}(p)}{p_{4}}\right)
  \nonumber\\&& \qquad \quad +  
  \mu-|\vec{p}|+\mathrm{Re}\Sigma^{+}_{+}(p)
  \Biggr)^{-1}  
\end{eqnarray}
near the Fermi surface has been subject of several weak coupling
analyses~\cite{Manuel:2000mk,Schafer:2004zf}. Especially the imaginary part of
the self energy, encoding the wave-function renormalization on the Fermi
surface has been studied in detail. Within the present truncation one is able
to generalize the investigation described in ref.~\cite{Schafer:2004zf} such
that the momentum dependence of the wave-function renormalization
is taken into account. For bare quarks, corresponding to a 1-loop
approximation, the contribution of the transversal gluons to the self energy
is given with $l=|\vec{p}\,|-p_{F}$ by
\begin{eqnarray}
  \Sigma^{+}_{+}(p)
  \simeq
  -\frac{4i}{3\pi^{2}}\int\!\! dk_{4}\int\!\! dk &&
  \arctan\left(\frac{k}{p_{4}+k_{4}+il}\right)
  \nonumber \\ && 
  \frac{\alpha_{s}(k^{2}) k}{k^{2}+\frac{\pi}{2}m^{2}(k^{2})\frac{|k_{4}|}{k}}.
\end{eqnarray}
The integrand is discontinuous for $|l|<k$, and the leading non-analytic
contribution can be extracted from
\begin{eqnarray}
  i \frac{\mathrm{Im}\Sigma^{+}_{+}(p)}{p_{4}}
 &\simeq&
  \frac{d}{dp_{4}}\Sigma^{+}_{+}(p)
  \\  
  &\simeq&
  -\frac{4i}{3\pi}\int_{k>|l|}\!\! dk
  \frac{\alpha_{s}(k^{2}) k}{k^{2}+\frac{\pi}{2}m^{2}(k^{2})\frac{|p_{4}|}{k}}.
\nonumber 
\end{eqnarray}
The main contribution to the integral comes from scales of the order
$k\sim \left(\pi\,m^{2}(k^{2})|p_{4}|\right)^{1/3}$. Note that 
$k \rightarrow 0$ for $p_{4}\rightarrow 0$. Employing this implicitly given
scale one can approximate
\begin{eqnarray}
  \label{nanaunb}
  \frac{\mathrm{Im}\Sigma^{+}_{+}(p)}{p_{4}}
  &\simeq&
  \frac{4}{9}\frac{\alpha_{s}(k^{2})}{\pi}
  \ln\left(\frac{\Bigl| |\vec{p}\,|-p_{F}\Bigr|^{3}+
      \frac{\pi}{2}m^{2}(k^{2})|p_{4}|}{\Lambda_{UV}^{3}}\right)
  \nonumber \\    
\end{eqnarray}
where $\Lambda_{UV}$ is an ultraviolet cutoff. This demonstrates the
well-known fact that the long-range (static) gluon interaction renders quark
matter in the unbroken phase into a non-Fermi liquid. In the present context
it is obvious that this non-trivial feature depends on the infrared behavior
of the product of gluon propagator and quark-gluon vertex.

Here also a qualitative difference between the couplings displayed in 
Fig.~\ref{salpha} becomes important. Since the coupling $\alpha_{I}(q^{2})$
becomes almost constant for $q<0.5\mathrm{GeV}$, we can estimate that $k$
becomes negligibly small, {\it i.e.\/} $k\sim 0$,  if
$$|p_{4}|<\frac{(0.5\mathrm{GeV})^{3}}{\pi m^{2}(0)}\approx
\frac{10\mathrm{MeV}}{\mu^{2}[\mathrm{GeV}^{2}]}.$$
For the coupling $\alpha_{II}(q^{2})$ this is not possible because the
coupling vanishes in the infrared. One would need $|p_{4}|\ll
{(10\mathrm{MeV})}/{(\mu^{2}[\mathrm{GeV}^{2}])}$ to study this effect, which
is far beyond the scope of the present work.

\begin{figure}
{\hspace{-0.01cm}\includegraphics[width=8.51cm]{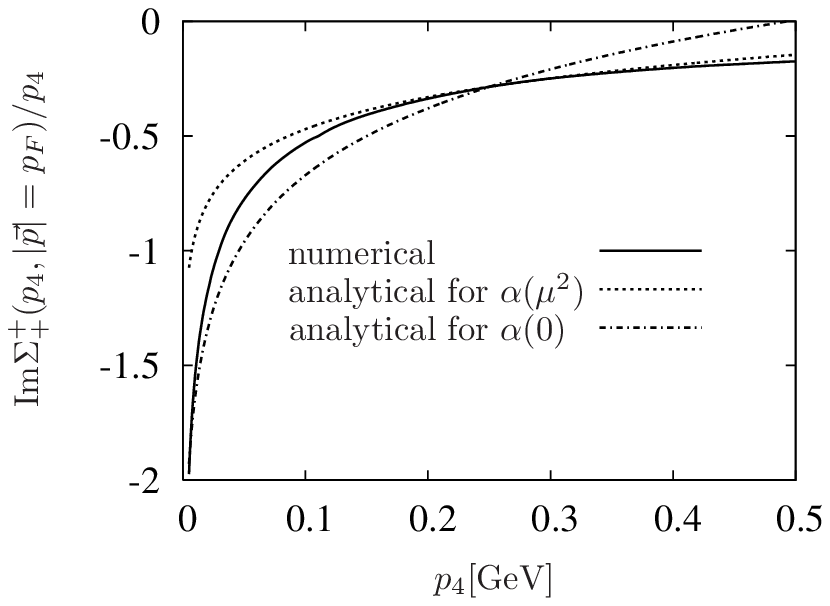}}
{\hspace{+0.cm}\includegraphics[width=8.5cm]{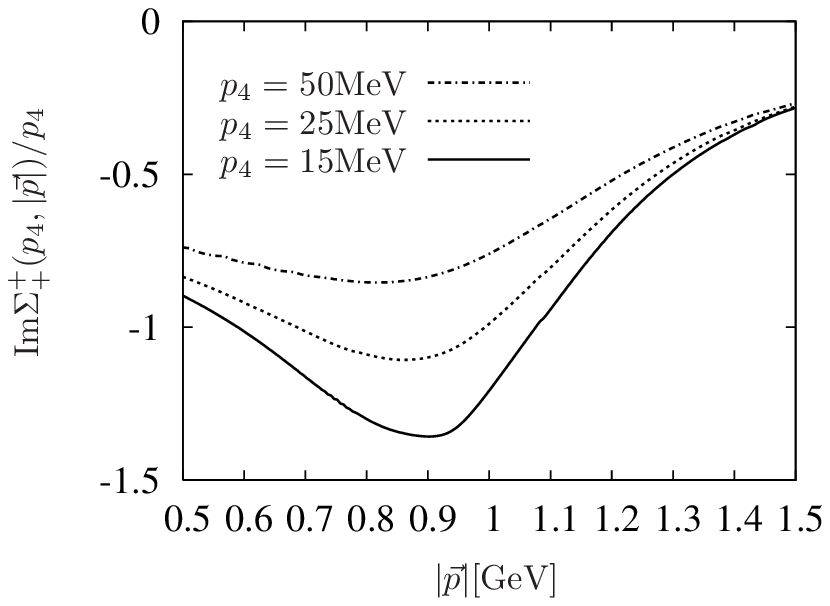}}
\caption{The 
  logarithmic singularity on the Fermi surface as a function of
  $p_{4}$, compared to the approximation given in eq.~(\ref{nanaunb}) at
  scales $k=0$ and $k=\mu$, (upper panel)
  and its momentum dependence for fixed $p_{4}\neq 0$ (lower panel).
  Here the coupling $\alpha_{I}(k^{2})$ is employed. $\Lambda_{UV}=1.3$GeV 
  is chosen such that the analytical approximation fits the numerical result.}
\label{imsigu}
\end{figure}

In Fig.~\ref{imsigu} we compare the non-analytic $p_{4}$ dependence at the
Fermi surface to the corresponding numerical result, employing the coupling 
$\alpha_{I}(k^{2})$. One clearly sees, that if $k$ and $\Lambda_{UV}$ are 
chosen accordingly, the approximation (\ref{nanaunb}) works well. In addition,
we display the dependence on $|\vec{p}|$ (which is neglected in weak coupling
analyses) demonstrating that the singularity only occurs at
$|\vec{p}|=p_{F}$.

\begin{figure}
{\hspace{-0.05cm}\includegraphics[width=8.51cm]{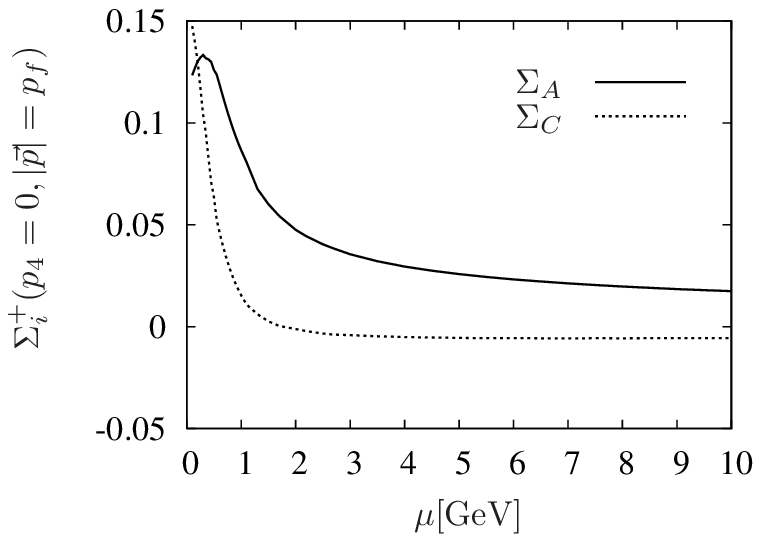}}
{\hspace{+0.05cm}\includegraphics[width=8.5cm]{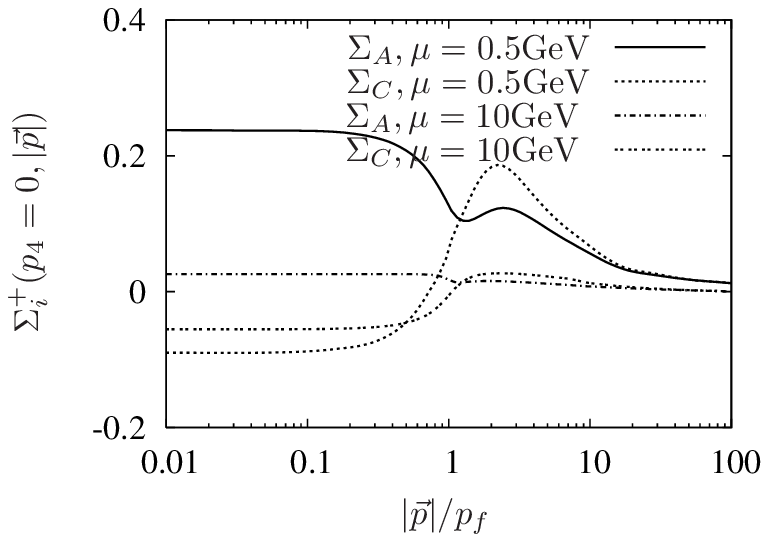}}
\caption{The self energies 
  $\Sigma_{A}^{+}$ and $\Sigma_{C}^{+}$ in the unbroken phase
  at the Fermi surface as a function of the chemical potential $\mu$ (upper
  panel) and as a function of
  the momentum $|\vec{p}|$ for
  $\mu=0.5\mathrm{GeV}$ and $10\mathrm{GeV}$ for $p_{4}=0$  (lower panel).
  Note that for $p_{4}=0$ all dressing functions are real.}
\label{resigu}
\end{figure}

For completeness, we present numerical results for the self-energy 
functions $\Sigma_{A}^{+}$ and $\Sigma_{C}^{+}$ in the unbroken phase, see 
Fig.~\ref{resigu}. Displayed are their values at the Fermi surface as a
function of the chemical potential $\mu$ and the momentum dependence for
$p_{4}=0$ at $\mu=0.5\mathrm{GeV}$ and $10\mathrm{GeV}$. One sees that, in the
unbroken phase, the approximation using bare quarks in the medium polarization
is justified {\it a posteriori} since, as expected,
$\Sigma_{A}^{+},\Sigma_{C}^{+}\ll 1$.

\section{Results for 2SC and CFL phase in the chiral limit}
In this section we combine the presentation of the numerical results for 
the scalar 2SC and CFL phases for three massless flavors. 

The color-flavor structure of the 2SC phase is given by
$M=\lambda_{2}\otimes\tau_{2}$ which leads to
\begin{eqnarray}
  S^{+\,-1}
  &=&
  S^{+\,-1}_{1} P_{ur,ug,dr,dg}+S^{+\,-1}_{2} P_{ub,db}+S^{+\,-1}_{3} P_{s},
  \nonumber \\
  \Phi^{+}
  &=&
  \frac{Z_{2}}{Z_{1F}} \, \phi^{+}_{2SC}\, \lambda_{2}\otimes\tau_{2}.
\end{eqnarray}
Here $(r,g,b)$ and $(u,d,s)$ label color and flavor. The matrices 
$\lambda_{i}$ and $\tau_{i}$ are the Gell-Mann matrices of color and flavor
space, respectively, completed by
$\lambda_{0}=\tau_{0}=\sqrt{\frac{2}{3}}\mathds{1}$. As described above,
$P_{i}$, denote projectors on the coupled color-flavor space.

The CFL phase is given by the ansatz
$M=\lambda_{2}\otimes\tau_{2}+\lambda_{5}\otimes\tau_{5}+
\lambda_{7}\otimes\tau_{7}$. Note that $M$ is antisymmetric in color and
flavor, respectively. The corresponding condensates can be chosen to be
invariant under a $3\otimes\bar{3}=1\oplus 8$ transformation with
generators $\tau_{a}-\lambda_{a}^{T}$~\cite{Alford:1998mk}. Then the matrices
$\{P_{i}\}$ turn out to project onto the irreducible representations. We
therefore obtain
\begin{eqnarray}
  S^{+\,-1}
  &=&
  S^{+\,-1}_{1} P_{1} + S^{+\,-1}_{8} P_{8},\\
  \Phi^{+} 
  &=&
  \frac{Z_{2}}{Z_{1F}}
  \left(
    \phi^{+}_{1}M_{1}+
    \phi^{+}_{8}M_{8}
  \right)\\
  &=&
  \frac{Z_{2}}{Z_{1F}}
  \left(
    \phi^{+}_{\bar{3}}\sum_{A=\{2,5,7\}}\,\lambda_{A}\otimes\tau_{A}
   \right.
   \nonumber\\ && \qquad \left. + \, 
    \phi^{+}_{6}\sum_{S=\{0,1,3,4,6,7,8\}}\,\lambda_{S}\otimes\tau_{S}
  \right), \nonumber
\end{eqnarray}
where we have also introduced the commonly used 'antitriplet' and 'sextet'
pairing function
$\phi^{+}_{\bar{3}}=\frac{1}{3}\phi_{1}+\frac{2}{3}\phi_{8}$ and
$\phi^{+}_{6}=-\frac{1}{3}\phi_{1}+\frac{1}{3}\phi_{8}$.

Both phases have been studied in a weak coupling analyses in the HDL
approximation. Including in this approximation the ``normal'' quark 
self energies, the quasiparticle gap at the Fermi surface is given
by~\cite{Wang:2001aq}
\begin{eqnarray}
  \phi^{+}_{weak, i}
  &=& 512\,\pi^{4}\left(\frac{2}{N_{f}g^{2}}\right)^{\frac{5}{2}}
  e^{-\frac{\pi^{2}+4}{8}} \,\mu\,
  e^{-\frac{3\pi^{2}}{\sqrt{2}g}}
  \nonumber \\ &&\times
  \left\{
    {1 \atop 2^{-1/3}}
    {i=\mathrm{2SC} \atop i=\bar{3}}
  \right. .
\end{eqnarray}
The momentum dependence takes a more complicated form but for
\mbox{$g^{2}\ln\left(\mu/\phi^{+}_{weak, i}\right)\ll1$}
one can neglect the quark self energies and obtains~\cite{Wang:2001aq}
\begin{eqnarray}
  \phi^{+}_{weak, i}\left(\lvert\vec{p}\rvert\right)
  &=&
  \phi^{+}_{weak, i}
  \\ &&
  \cos\left(
    \frac{g}{3\sqrt{2}\pi}\ln
    \left(
      \frac{\phi^{+}_{weak, i}}{|p-\mu|+\epsilon_{i}^{+}(p)}
    \right)
  \right),
  \nonumber 
  \label{weakgapp}
\end{eqnarray}
with $\epsilon_{i}^{+}(p)^{2}=(p-\mu)^{2}+{\phi^{+}_{weak, i}}^{2}$.

For the weak-coupling approximations of the gap functions displayed in the
following figures we use the identical running coupling as the one employed
to obtain the displayed numerical solution, respectively. Note, however, that
in the strong coupling regime the weak-coupling expressions ceases to be valid.

\begin{figure}
{\hspace{+0.05cm}\includegraphics[width=8.5cm]{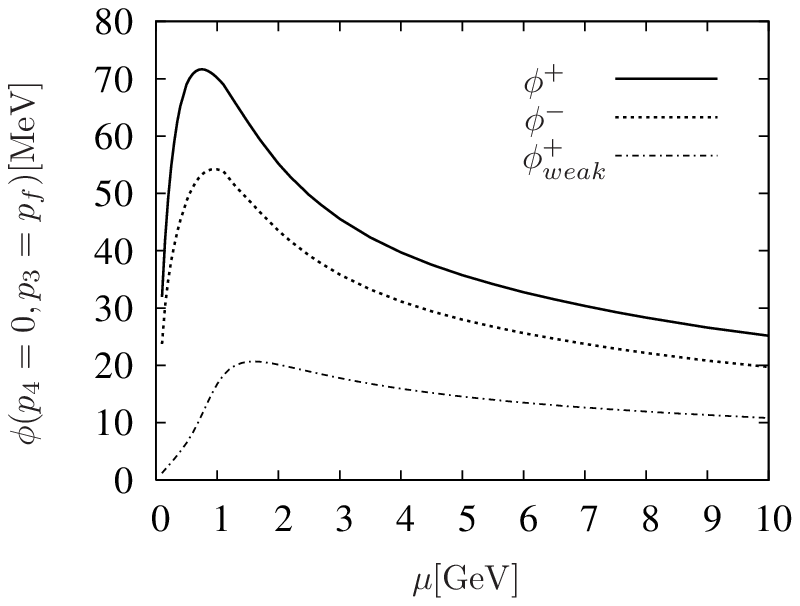}}
{\hspace{+0.05cm}\includegraphics[width=8.5cm]{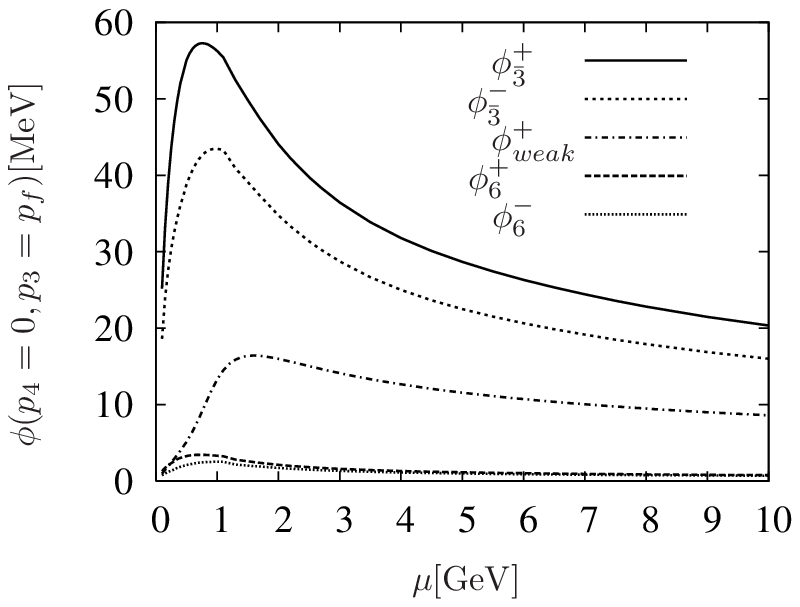}}
\caption{The quasiparticle gaps $\phi^{+}$ and 
  anti-quasiparticle gaps
  $\phi^{-}$ at the Fermi surface for the 2SC (upper panel) and CFL 
  (lower panel) phases for the coupling $\alpha_{I}(k^{2})$.
  These are compared to the extrapolated weak coupling result
  $\phi^{+}_{weak}$.}
\label{Dpapsmu}
\end{figure}
\begin{figure}
{\hspace{+0.05cm}\includegraphics[width=8.5cm]{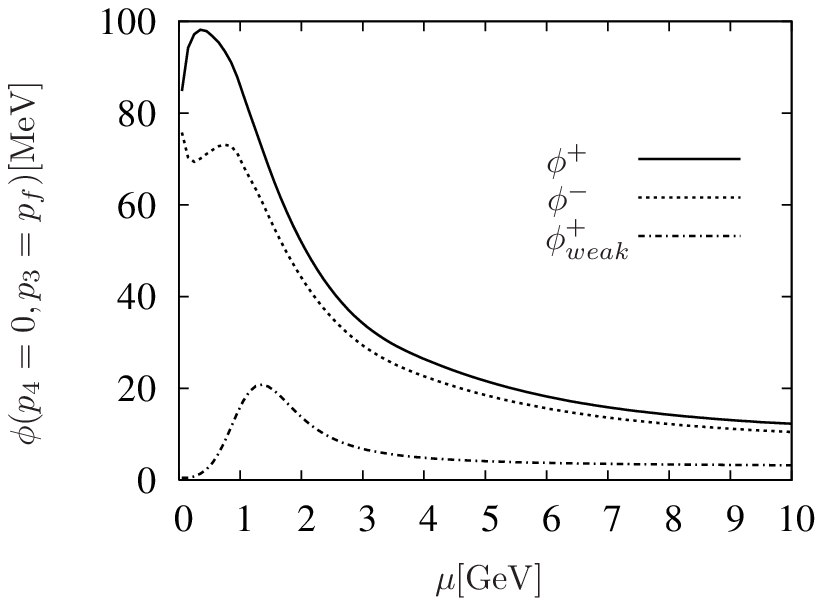}}
{\hspace{+0.05cm}\includegraphics[width=8.5cm]{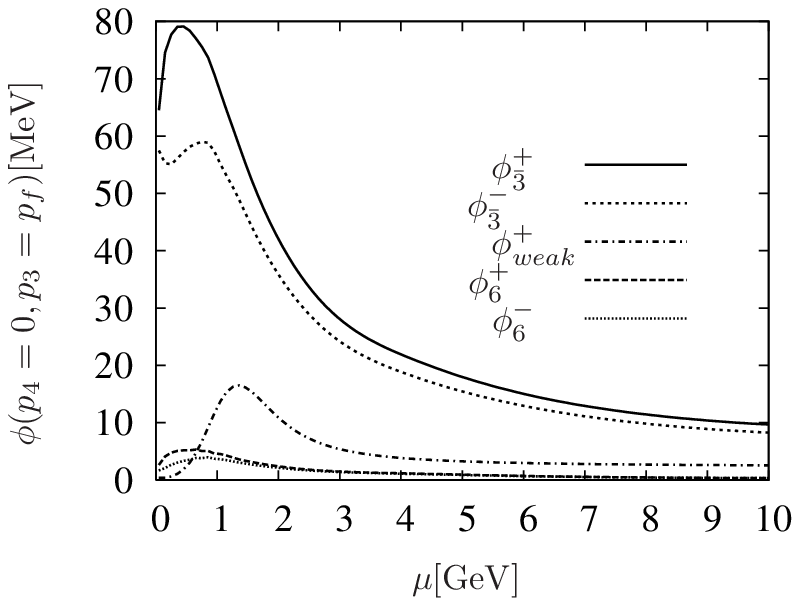}}
\caption{The quasiparticle gaps $\phi^{+}$ and 
  anti-quasiparticle gaps
  $\phi^{-}$ at the Fermi surface for the 2SC (upper panel) and CFL 
  (lower panel) phases for
  coupling $\alpha_{II}(k^{2})$. These are compared to the 
  extrapolated weak coupling
  result $\phi^{+}_{weak}$.}
\label{lpapsmu}
\end{figure}
\begin{figure}
{\hspace{-0.05cm}\includegraphics[width=8.5cm]{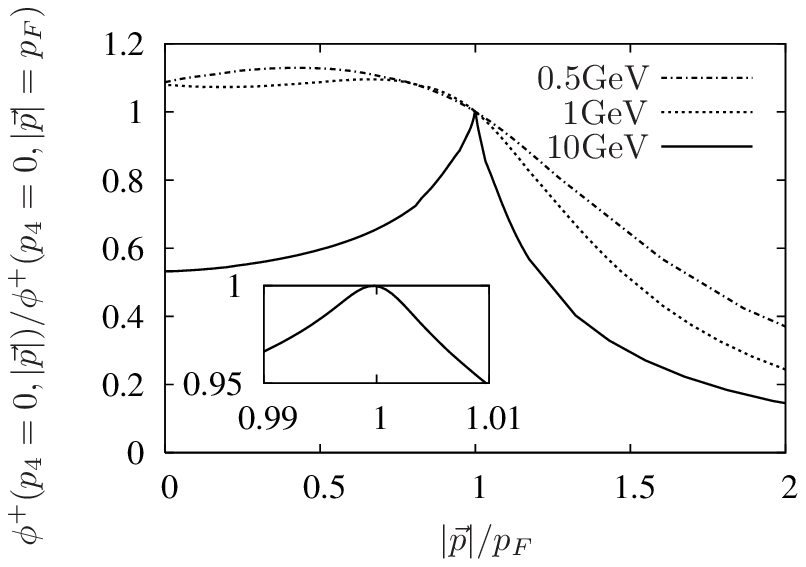}}
{\hspace{+0.05cm}\includegraphics[width=8.5cm]{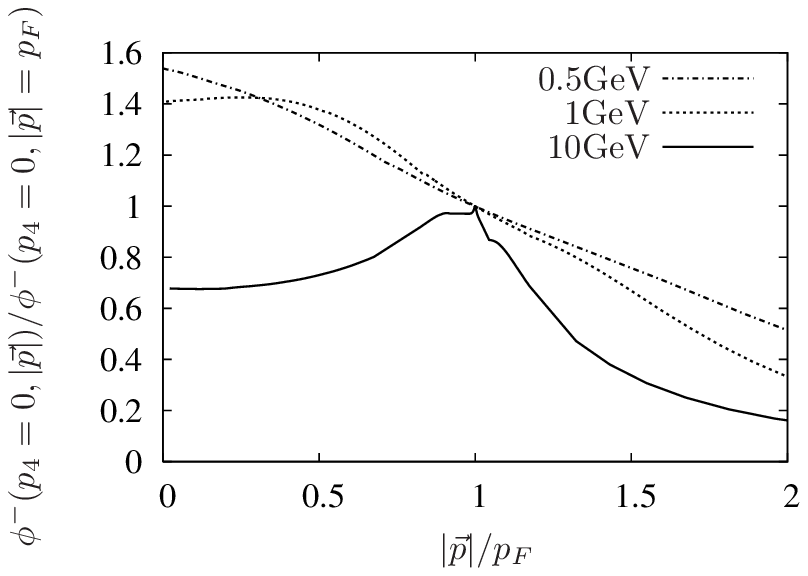}}
\caption{The momentum dependence of the quasiparticle gaps 
  $\phi^{+}$ (upper panel) and
  anti-quasiparticle gaps $\phi^{-}$ (lower panel) for $p_{4}=0$ 
  normalized to their
  value at the Fermi surface.}
\label{Dpapsp3}
\end{figure}

The results for the gap functions at the Fermi surface the for 2SC and CFL
phases are shown in Fig.~\ref{Dpapsmu} for the coupling
$\alpha_{I}(k^{2})$. In
Fig.~\ref{lpapsmu} the corresponding results are shown for the coupling
$\alpha_{II}(k^{2})$.  In both figures it is evident that the
extrapolated weak-coupling results for the value of the quasiparticle gap
differs by a factor of more than five for $\mu\approx 500\mathrm{MeV}$,
resulting in quasiparticle-pairing gaps larger than $50\mathrm{MeV}$. It is
worth noting that the results are less sensitive on the coupling than the mass
function in the chirally broken phase. This can be attributed to the fact that 
a stronger coupling also results in stronger screening and damping.  In
addition, also the anti-quasiparticle gap function are determined
self-consistently. Nevertheless, it is remarkable that the ratio of the 2SC
and anti-triplet CFL gaps is almost equal to $2^{1/3}$ as in the weak coupling
analysis. The reason for this behavior is that the normal
self energies $\Sigma^{+}_{i}$ are only weakly modified compared to the
self energies in the unbroken phase. This amounts to an effective
``decoupling'' of normal self energies and gap functions, and thus the factor
$2^{1/3}$ is directly inherited from the pairing pattern in color-flavor space.

As can be inferred from Fig.~\ref{resigu} and the estimate given in 
eq.~(\ref{DeltaE}), the energy gap in the excitation spectrum $\Delta_{i}^{e}$
is at most $15\%$ smaller than the gap function at the corresponding
kinematical point. (Note that the corresponding coefficients $\delta_i$ in
the decomposition of $M^{\dagger}M$ are given by:
$\sqrt{\delta_{2SC}}=\sqrt{\delta_{1}}=1$ and $\sqrt{\delta_{8}}=2$.)
We also emphasize, that $\phi^{+}(p)\rightarrow\phi^{-}(p)$ for
$\mu\rightarrow 0$, which can be non-zero, indicating a Bose-Einstein
condensation of diquarks. This is found for the coupling
$\alpha_{II}(k^{2})$. However, this is not expected to be the energetically
favored state in the vacuum, since spontaneous chiral symmetry breaking has not
been taken into account here.

In Fig.~\ref{Dpapsp3} we present the momentum dependence of the quasiparticle
gaps $\phi^{+}$ and anti-quasiparticle gaps $\phi^{-}$ for $p_{4}=0$ for
several chemical potentials, obtained with the coupling
$\alpha_{I}(k^{2})$. For large values of $\mu$ the gap function is
concentrated around the Fermi
surface and the quasiparticle gaps show a cosine-like behavior near the
maximum. However, at chemical potentials of the order $\Lambda_{QCD}$, the
Fermi surface is no longer the main contributing region to the gap integral,
and even the maximum of the gap function is no longer on or close to the Fermi
surface.

\begin{figure}
{\hspace{+0.05cm}\includegraphics[width=8.5cm]{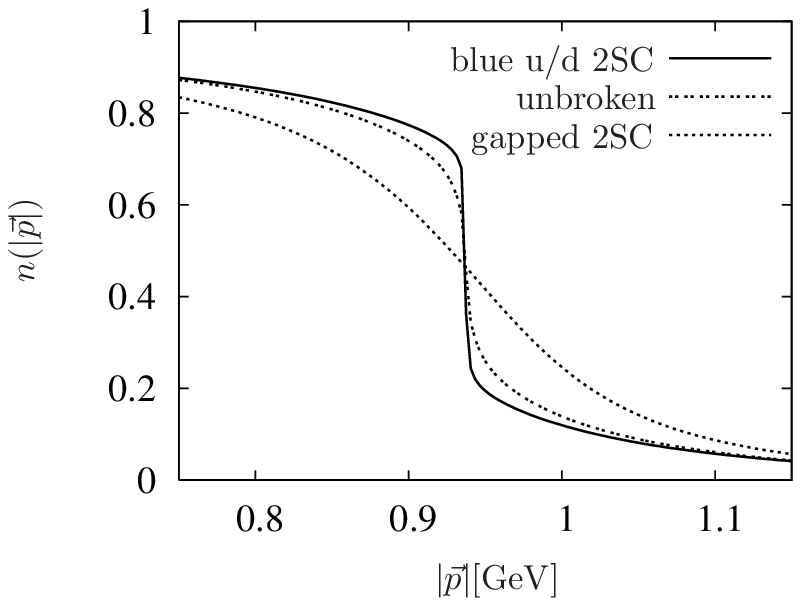}}
{\hspace{+0.05cm}\includegraphics[width=8.5cm]{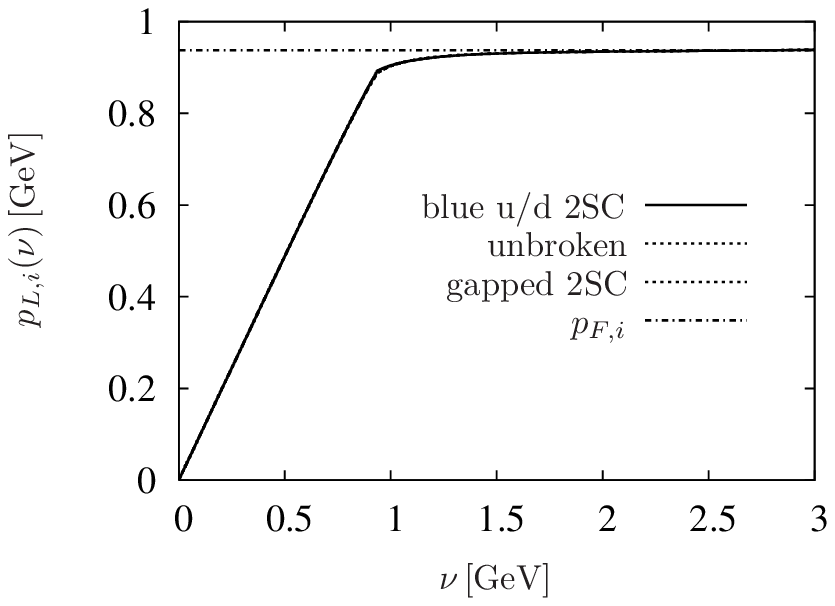}}
\caption{The 
  occupation numbers (upper panel) and the function defined in eq.\
  (\ref{pL}), $p_{L,i}(\nu)$,  (lower panel)
  for the different quark channels in the 2SC phase at
  $\mu=1\mathrm{GeV}$ with the coupling $\alpha_{I}(k^{2})$. 
  As can be seen in the panels all Fermi momenta are (almost) equal 
  and given by 
  $p_{F,i}\approx .937\mathrm{GeV}$, {\it c.f.\/} the dashed-dotted line
  in the lower panel. The lines of the different channels are nearly
  indistinguishable.}
\label{np3}
\end{figure}

As described in sect.~\ref{occcor}, the occupation numbers of the
quasiparticles are calculated from the normal propagators. These are displayed
in the upper panel of Fig.~\ref{np3} for the 2SC phase, employing the coupling
$\alpha_{I}(k^{2})$. The occupation number of the gapped red and green $u$- and
$d$-quarks is, as expected and due to the pairing, a smooth function around the
Fermi momentum. On the other hand, the occupation number of the blue $u$- and
$d$-quarks changes rapidly. Within the present approximation, {\it i.e.\/}
neglecting quark self energies in the medium polarization, these quarks are 
in a state at the borderline between a Fermi and a non-Fermi liquid. 

The decoupled $s$-quarks are in the unbroken phase and one clearly sees,
especially when comparing to the blue $u$- and $d$-quarks, that the occupation
number changes smoothly. As already mentioned, this is another feature of
non-Fermi liquids which, strictly speaking, indicates the breakdown of the
quasiparticle picture. We also find a significant depletion at
$\vert\vec{p}\vert=0$ due to the interaction.

The density can either be obtained by integration of the occupation numbers or
via the Luttinger theorem (see sect.~\ref{app2}). The conditions for the
latter are fulfilled since all self-energies are real for $p_{4}=0$.
To analyze this remarkable feature we define the function
\begin{eqnarray}
  p_{L,i}(\nu ) &=& \left(3\int_{0}^{\nu }dq q^{2}
  n_{i}(q)\right)^{\frac{1}{3}},
\label{pL}
\end{eqnarray}
which has to obey the limiting behavior 
$\lim_{\nu \rightarrow\infty}p_{L,i}(\nu )=p_{F,i}$.
From Fig.~\ref{np3} one sees that this limit is assumed for $\nu>p_{F,i}$, and 
that the expectation derived from Luttinger's theorem is nicely fulfilled 
within the numerical accuracy \footnote{We thank D.T. Son for corresponding remarks which allowed us to perform this check on our results.}. 
Note also that all Fermi momenta are very close to each other, a fact which 
justifies {\it a posteriori} the approximation of neglecting the neutrality 
conditions.

\begin{figure}
{\hspace{0.cm}\includegraphics[width=8.5cm]{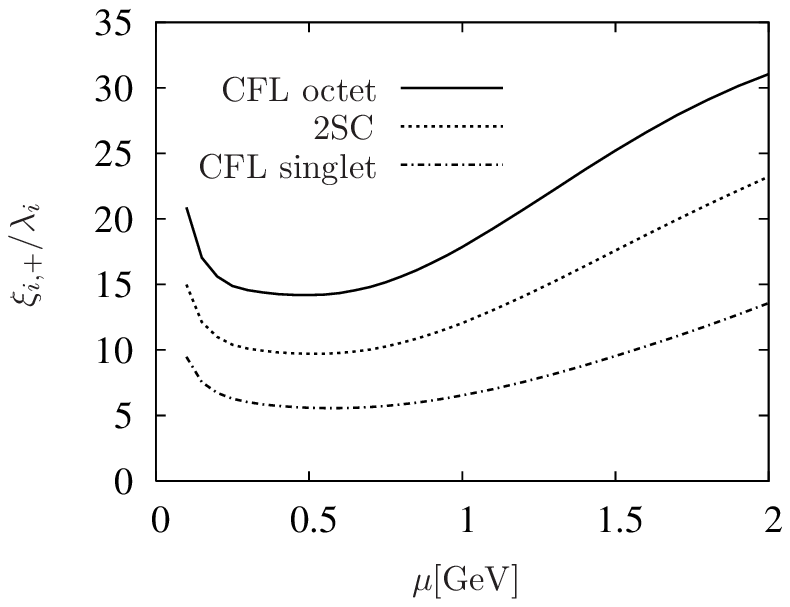}}
{\hspace{0.cm}\includegraphics[width=8.5cm]{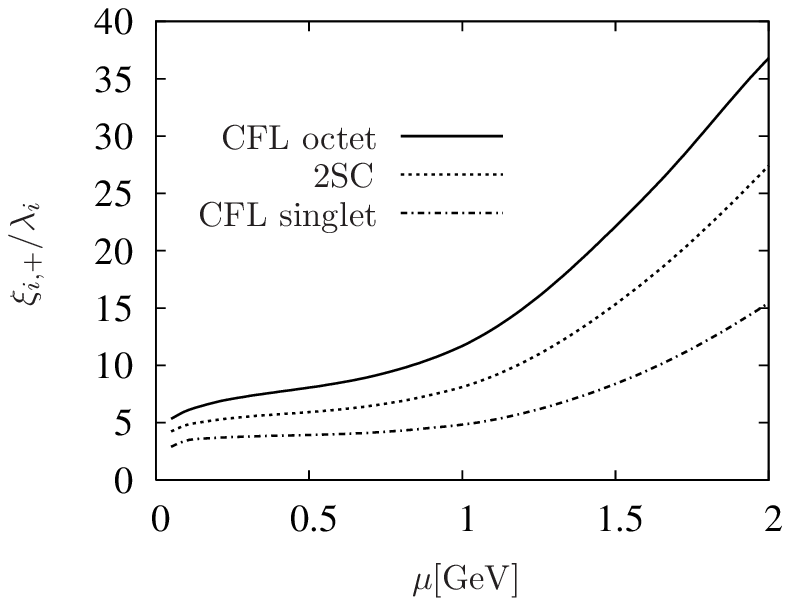}}
\caption{The ratio between the 
  coherence length and mean-free path in 2SC and CFL phase for the coupling 
  $\alpha_{I}(k^{2})$ (upper panel) and $\alpha_{II}(k^{2})$ (lower panel).}
\label{corlen}
\end{figure}

It is instructive to compare the coherence length of the diquarks, $\xi_{+}$, 
to the mean-free path as determined from the density. The
results are shown in Fig.~\ref{corlen} as a function of the chemical potential.
Although the two different couplings lead to a distinctive pattern for these
ratios it is safe to conclude that the size of a Cooper pair at 
moderate chemical potentials, $\mu\approx 500\mathrm{MeV}$, is only
several times the mean-free path, the precise value depending on the diquark
channel and the employed coupling (similar results were already presented in
ref.~\cite{Abuki:2001be} for the 2SC phase.). Although there is an analogy to
the crossover between a BCS-type superconductor in weak coupling to the
strongly coupled regime (Bose-Einstein condensate) the size of
the diquarks suggest that the 2SC and CFL phases resemble a strongly coupled
BCS system. However the result indicates, that mean-field approximations are
very questionable. In the limit $\mu\rightarrow 0$, we again find a different
behavior for the two employed couplings indicating Bose-Einstein condensation
of diquarks for $\alpha_{II}(k^{2})$ for very small chemical potentials.

\begin{figure}
{\hspace{0.cm}\includegraphics[width=8.5cm]{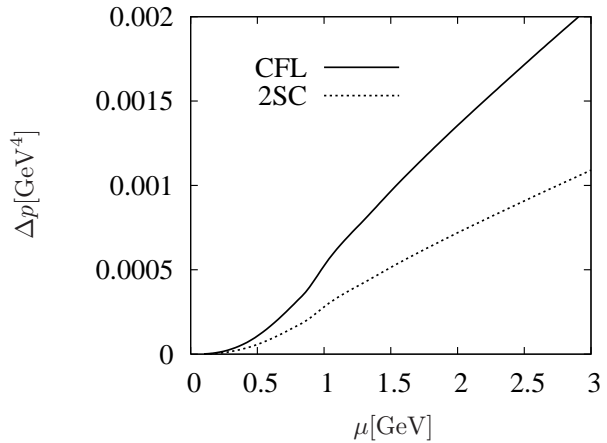}}
\caption{The pressure difference between the 
  superconducting phases (2SC and CFL) 
  and the unbroken phase employing $\alpha_{I}(k^{2})$.}
\label{m0CJT}
\end{figure}

Finally, the pressure difference between the superconducting phases 
(2SC and CFL)  and the unbroken phase are displayed in Fig.~\ref{m0CJT}. 
This quantity is given by the negative effective action difference, and is
here calculated within the approximate CJT formula given in eq.\
(\ref{CJT}). In the chiral limit, the CFL phase is the most favored phase, as
expected. This is in good qualitative agreement with the parameterized form
\begin{eqnarray}
  \Delta p
  &=& 
  \sum_{i} \mathrm{rank}(P_{i})\,
  \frac{p_{F,i}^{2}\,\Delta^{+\,2}_{i}}{8\pi^{2}}\,.
\end{eqnarray}
which is (slightly) a generalized formula as compared to the one obtained 
in the weak coupling regime~\cite{Schmitt:2004et}.

\section{Conclusions and outlook}

We have presented the solution of a truncated Dyson-Schwinger equation for the
quark propagator at non-vanishing chemical potential in the chiral limit. The
truncation includes the in-medium modification of the gluon propagator and has
been chosen such that, at vanishing density, known results for the vacuum and
at asymptotically high densities the corresponding weak-coupling expressions
are recovered. We have determined analytically the ultraviolet behavior, {\it
  i.e.\/} the anomalous dimensions, of the gap functions.

The most remarkable feature of these self-consistent solutions are the values
of the gap functions: The quasiparticle-pairing gaps are several times larger
than the extrapolated weak-coupling results even for sizeable chemical
potentials $\mu \approx 1$GeV. This is true irrespective of the considered
phase (2SC or CFL) and the employed running coupling.

The investigation, presented here, provides the starting point for more
realistic studies involving finite current masses. Using different quark
masses the question of implementing neutrality conditions (which could be
ignored here) becomes important.

Due to solutions of the gluon- and ghost Dyson-Schwinger equations at
non-vanishing temperatures~\cite{Maas} an investigation of different
regions of the phase diagram opens up. However, to solve the complete coupled
system of the Dyson-Schwinger equations for the QCD propagators remains a 
challenge. A study of the question of deconfinement due to a non-vanishing
quark density
requires certainly also a better understanding of the infrared behavior of
these Green's functions and thus an infrared analysis of the coupled
Dyson-Schwinger system at non-vanishing temperatures and densities.

\section*{Acknowledgments}
We thank Dirk Rischke and Igor Shovkovy for helpful discussions, Michael
Buballa, Axel Maas, and Kai Schwenzer for helpful discussions as well as a
critical reading of the manuscript. D.N.\ is grateful to the DAAD for a "DAAD Doktorandenstipendium", and to the members of the FWF-funded  Doctoral Program ``Hadrons in vacuum, nuclei and stars'' at the Institute of Physics of the University of Graz for their warm hospitality. 

This work has been furthermore supported in part by the Helmholtz association
(Virtual Theory Institute VH-VI-041) and by the BMBF under grant number
06DA916.


\begin{thebibliography}{99}
%
\bibitem{reviews}
K.~Rajagopal and F.~Wilczek,
arXiv:hep-ph/0011333;
T.~Schafer,
arXiv:hep-ph/0304281;
M.~Buballa,
Phys.\ Rept.\  {\bf 407}, 205 (2005)
[arXiv:hep-ph/0402234];
I.~A.~Shovkovy,
arXiv:nucl-th/0511014.

%
\bibitem{Rischke:2003mt}
D.~H.~Rischke,
Prog.\ Part.\ Nucl.\ Phys.\  {\bf 52}, 197 (2004)
[arXiv:nucl-th/0305030].

%
\bibitem{SchafHong}
D.~T.~Son,
Phys.\ Rev.\ D {\bf 59}, 094019 (1999)
[arXiv:hep-ph/9812287].
T.~Schafer and F.~Wilczek,
Phys.\ Rev.\ D {\bf 60}, 114033 (1999)
[arXiv:hep-ph/9906512]; 
D.~K.~Hong, V.~A.~Miransky, I.~A.~Shovkovy and L.~C.~R.~Wijewardhana,
Phys.\ Rev.\ D {\bf 61}, 056001 (2000)
[Erratum-ibid.\ D {\bf 62}, 059903 (2000)]
[arXiv:hep-ph/9906478];
I.~A.~Shovkovy and L.~C.~R.~Wijewardhana,
Phys.\ Lett.\ B {\bf 470}, 189 (1999)
[arXiv:hep-ph/9910225].

%
\bibitem{AlfRapp}
M.~G.~Alford, K.~Rajagopal and F.~Wilczek,
Phys.\ Lett.\ B {\bf 422}, 247 (1998)
[arXiv:hep-ph/9711395];
R.~Rapp, T.~Schafer, E.~V.~Shuryak and M.~Velkovsky,
Phys.\ Rev.\ Lett.\  {\bf 81}, 53 (1998)
[arXiv:hep-ph/9711396].

%
\bibitem{Greensite:2003bk}
J.~Greensite,
Prog.\ Part.\ Nucl.\ Phys.\  {\bf 51}, 1 (2003)
[arXiv:hep-lat/0301023].

%
\bibitem{Greensite:2004mh}
J.~Greensite, S.~Olejnik and D.~Zwanziger,
AIP Conf.\ Proc.\  {\bf 756}, 162 (2005)
[arXiv:hep-lat/0411032].

%
\bibitem{Zwanziger:2003de}
D.~Zwanziger,
Phys.\ Rev.\ D {\bf 70}, 094034 (2004)
[arXiv:hep-ph/0312254].

%
\bibitem{vonSmekal:2000pz}
L.~von Smekal and R.~Alkofer,
in: Proceedings of the Fourth International Conference on Quark
Confinement and the Hadron Spectrum (CONFINEMENT IV), July 3-8, Vienna
[arXiv:hep-ph/0009219].

%
\bibitem{Zwanziger:2002sh}
D.~Zwanziger,
Phys.\ Rev.\ Lett.\  {\bf 90}, 102001 (2003)
[arXiv:hep-lat/0209105].

%
\bibitem{Nakamura:2005ux}
A.~Nakamura and T.~Saito,
arXiv:hep-lat/0512042.

%
\bibitem{Alkofer:2005ug}
R.~Alkofer, M.~Kloker, A.~Krassnigg and R.~F.~Wagenbrunn,
Phys.\ Rev.\ Lett.\  {\bf 96}, 022001 (2006)
[arXiv:hep-ph/0510028].


%
\bibitem{Silva:2005hb}
P.~J.~Silva and O.~Oliveira,
arXiv:hep-lat/0511043.

%
\bibitem{Sternbeck:2005tk}
A.~Sternbeck, E.~M.~Ilgenfritz, M.~Mueller-Preussker and A.~Schiller,
Phys.\ Rev.\ D {\bf 72}, 014507 (2005)
[arXiv:hep-lat/0506007].

%
\bibitem{Fischer:2004uk}
C.~S.~Fischer and H.~Gies,
JHEP {\bf 0410}, 048 (2004)
[arXiv:hep-ph/0408089].

%
\bibitem{Pawlowski:2003hq}
J.~M.~Pawlowski, D.~F.~Litim, S.~Nedelko and L.~von Smekal,
Phys.\ Rev.\ Lett.\  {\bf 93}, 152002 (2004)
[arXiv:hep-th/0312324].

%
\bibitem{Alkofer:2003jj}
R.~Alkofer, W.~Detmold, C.~S.~Fischer and P.~Maris,
Phys.\ Rev.\ D {\bf 70}, 014014 (2004)
[arXiv:hep-ph/0309077].

%
\bibitem{Maas}
A.~Maas, B.~Gruter, R.~Alkofer and J.~Wambach,
arXiv:hep-ph/0210178;
A.~Maas, J.~Wambach, B.~Gruter and R.~Alkofer,
Eur.\ Phys.\ J.\ C {\bf 37}, 335 (2004)
[arXiv:hep-ph/0408074];
B.~Gruter, R.~Alkofer, A.~Maas and J.~Wambach,
Eur.\ Phys.\ J.\ C {\bf 42}, 109 (2005)
[arXiv:hep-ph/0408282];
A.~Maas, J.~Wambach and R.~Alkofer,
Eur.\ Phys.\ J.\ C {\bf 42}, 93 (2005)
[arXiv:hep-ph/0504019];
A.~Maas,
Mod.\ Phys.\ Lett.\ A {\bf 20}, 1797 (2005)
[arXiv:hep-ph/0506066].
  
%
\bibitem{Fischer:2003rp}
C.~S.~Fischer and R.~Alkofer,
Phys.\ Rev.\ D {\bf 67}, 094020 (2003)
[arXiv:hep-ph/0301094].

%
\bibitem{Wang:2001aq}
Q.~Wang and D.~H.~Rischke,
Phys.\ Rev.\ D {\bf 65}, 054005 (2002)
[arXiv:nucl-th/0110016].

%
\bibitem{Gorkov}
L.\ P.\ Gorkov, Sov.\ Phys.\ JETP {\bf 9} (1959) 1364;
Y.\ Nambu, Phys.\ Rev.\ {\bf 117} (1960) 648.

%
\bibitem{Alkofer:2000wg}
R.~Alkofer and L.~von Smekal,
Phys.\ Rept.\  {\bf 353}, 281 (2001)
[arXiv:hep-ph/0007355].
%
\bibitem{Roberts:2000aa}
C.~D.~Roberts and S.~M.~Schmidt,
Prog.\ Part.\ Nucl.\ Phys.\  {\bf 45}, S1 (2000)
[arXiv:nucl-th/0005064].

%
\bibitem{Rischke:2000ra}
D.~H.~Rischke,
Phys.\ Rev.\ D {\bf 62}, 054017 (2000)
[arXiv:nucl-th/0003063].

%
\bibitem{Gerhold:2003js}
A.~Gerhold and A.~Rebhan,
Phys.\ Rev.\ D {\bf 68}, 011502 (2003)
[arXiv:hep-ph/0305108].

%
\bibitem{Dietrich:2003nu}
D.~D.~Dietrich and D.~H.~Rischke,
Prog.\ Part.\ Nucl.\ Phys.\  {\bf 53}, 305 (2004)
[arXiv:nucl-th/0312044].

%
\bibitem{Buballa:2005bv}
M.~Buballa and I.~A.~Shovkovy,
Phys.\ Rev.\ D {\bf 72}, 097501 (2005)
[arXiv:hep-ph/0508197].

%
\bibitem{Fischer:2002hn}
C.~S.~Fischer and R.~Alkofer,
Phys.\ Lett.\ B {\bf 536}, 177 (2002)
[arXiv:hep-ph/0202202];
C.~S.~Fischer, B.~Gruter and R.~Alkofer,
Ann. Phys. (2006), in print [arXiv:hep-ph/0506053].

%
\bibitem{Bowman:2004jm}
P.~O.~Bowman, U.~M.~Heller, D.~B.~Leinweber, M.~B.~Parappilly and A.~G.~Williams,
Phys.\ Rev.\ D {\bf 70}, 034509 (2004)
[arXiv:hep-lat/0402032].

%
\bibitem{Llanes-Estrada:2004jz}
F.~J.~Llanes-Estrada, C.~S.~Fischer and R.~Alkofer,
arXiv:hep-ph/0407332.

%
\bibitem{Skullerud:2004gp}
J.~I.~Skullerud, P.~O.~Bowman, A.~Kizilersu, D.~B.~Leinweber and A.~G.~Williams,
Nucl.\ Phys.\ Proc.\ Suppl.\  {\bf 141}, 244 (2005)
[arXiv:hep-lat/0408032].

%
\bibitem{Lin:2005zd}
H.~W.~Lin,
arXiv:hep-lat/0510110.

%
\bibitem{Bhagwat:2003vw}
M.~S.~Bhagwat, M.~A.~Pichowsky, C.~D.~Roberts and P.~C.~Tandy,
Phys.\ Rev.\ C {\bf 68}, 015203 (2003)
[arXiv:nucl-th/0304003].

%
\bibitem{Fischer:2005nf}
C.~S.~Fischer and M.~R.~Pennington,
Phys.\ Rev.\ D {\bf 73}, 034029 (2006)
arXiv:hep-ph/0512233.

%
\bibitem{Kapusta:1989tk}
J.~I.~Kapusta, {\em Finite-temperature field theory} 
(Cambridge University Press, Cambridge, 1993).

%
\bibitem{Manuel:2000nh}
C.~Manuel,
Phys.\ Rev.\ D {\bf 62}, 114008 (2000)
[arXiv:hep-ph/0006106].

%
\bibitem{Pisarski:1999av}
R.~D.~Pisarski and D.~H.~Rischke,
Phys.\ Rev.\ D {\bf 60}, 094013 (1999)
[arXiv:nucl-th/9903023].
%

\bibitem{Gusynin:1986fu}
V.~P.~Gusynin and V.~A.~Miransky,
Phys.\ Lett.\ B {\bf 191}, 141 (1987).

%
\bibitem{Roberts:1994dr}
C.~D.~Roberts and A.~G.~Williams,
Prog.\ Part.\ Nucl.\ Phys.\  {\bf 33}, 477 (1994)
[arXiv:hep-ph/9403224].

%
\bibitem{Cornwall:1974vz}
J.~M.~Cornwall, R.~Jackiw and E.~Tomboulis,
Phys.\ Rev.\ D {\bf 10}, 2428 (1974).

%
\bibitem{Schmitt:2004et}
A.~Schmitt,
Phys.\ Rev.\ D {\bf 71}, 054016 (2005)
[arXiv:nucl-th/0412033].

%
\bibitem{Manuel:2000mk}
C.~Manuel,
Phys.\ Rev.\ D {\bf 62}, 076009 (2000)
[arXiv:hep-ph/0005040].

%
\bibitem{Schafer:2004zf}
T.~Schafer and K.~Schwenzer,
Phys.\ Rev.\ D {\bf 70}, 054007 (2004)
[arXiv:hep-ph/0405053].

%
\bibitem{Alford:1998mk}
M.~G.~Alford, K.~Rajagopal and F.~Wilczek,
Nucl.\ Phys.\ B {\bf 537}, 443 (1999)
[arXiv:hep-ph/9804403].

%
\bibitem{Abuki:2001be}
H.~Abuki, T.~Hatsuda and K.~Itakura,
Phys.\ Rev.\ D {\bf 65}, 074014 (2002)
[arXiv:hep-ph/0109013].

\end{thebibliography}
\end{document}